\documentclass{article}
\usepackage[utf8]{inputenc}
\usepackage[margin=1in,top=15 mm]{geometry}
\usepackage{authblk}
\usepackage{setspace}
\usepackage[colorlinks,citecolor=blue,urlcolor=blue,bookmarks=false,hypertexnames=true]{hyperref} 
\usepackage{graphicx}
\usepackage{multirow}
\usepackage{verbatim}
\usepackage{float}
\usepackage{hyperref}
\usepackage[nameinlink,noabbrev]{cleveref}
\usepackage[superscript,biblabel]{cite}

\begin{document}

\sloppy

\title{\textbf{Embracing Nonlinearity and Geometry: A dimensional analysis guided design of shock absorbing materials}}

\def\correspondingauthor{\footnote{Corresponding author: thevamaran@wisc.edu}}

\author[a]{Abhishek Gupta}
\author[a]{Komal Chawla}
\author[a]{Bhanugoban Maheswaran}
\author[a]{Daniyar Syrlybayev}
\author[a]{Ramathasan Thevamaran \correspondingauthor{}}

\affil[a]{Department of Mechanical Engineering, University of Wisconsin-Madison, Madison, WI, 53706, USA}

\maketitle

\vspace*{-\baselineskip}
\vspace*{-\baselineskip}
\par\noindent\rule{\textwidth}{0.5pt}
\smallskip

\begin{abstract}

\onehalfspacing

Protective applications require energy-absorbing materials that are soft and compressible enough to absorb kinetic energy from impacts, yet stiff enough to bear crushing loads. Achieving this balance requires careful consideration of both mechanical properties of the material and geometry of the shock-absorbing pads. Conventional shock-absorbing pads are typically made from very thick foams that exhibit a plateau of constant stress in their stress-strain response, while foams with a non-linearly stiffening stress-strain response are often considered ineffective. Contrary to this belief, we demonstrate that foams with a nonlinear stress-strain response can be effective for achieving protective pads that are both thin and lightweight, particularly for pad geometries requiring a large cross-sectional area. We introduce a new framework for the thickness or volume-constrained design of compact and lightweight protective foams while ensuring the desired structural integrity and mechanical performance. Our streamlined dimensional analysis provides geometric constraints on the dimensionless thickness and cross-sectional area of a protective foam with a given stress-strain response to limit the acceleration and compressive strain within desired critical limits. We also identify optimal mechanical properties that will result in the most compact and lightest protective foam pad for absorbing the given kinetic energy of impact. Guided by this design framework, we achieve optimal protective properties in hierarchically architected vertically aligned carbon nanotube (VACNT) foams, enabling next generation protective applications in extreme environments.


\end{abstract}

\doublespacing

\section*{Introduction}

Energy-absorbing materials permeate our lives, from soft polymeric foams used in helmet liners \cite{ramirez2018evaluation,shuaeib2007new}, packaging \cite{zhang1994mechanical}, and seat cushions, to crushable metallic foams employed in ballistic impact attenuators \cite{li2007dynamic,barnes2014dynamic}, automotive buffers, and planetary landers \cite{li2011crashworthiness}. Those protective foams must absorb the kinetic energy from impacts and undesirable vibrations while limiting the forces and accelerations imparted on the protected objects \cite{gibson2003cellular,zhang1994mechanical}. Compared to stochastic foams, architected foams demonstrate superior modulus, strength, and energy absorption at comparable or lower densities \cite{schaedler2016architected}. This indicates better specific (density-normalized) properties, achieved through architectural design of the lattice unit cells \cite{bauer2016approaching,chen2019stiff,crook2020plate,zheng2023unifying}. The pursuit of achieving specific mechanical properties near the theoretical limits has resulted in several advancements \cite{berger2017mechanical,bauer2016approaching}, including ultra-stiff micro-lattices \cite{schaedler2011ultralight,zheng2014ultralight}, nanolattices with high mechanical strength \cite{bauer2016approaching,wang2022achieving}, high energy-absorption of supersonic projectiles \cite{portela2021supersonic,butruille2024decoupling}, and fracture resilience in hierarchical \cite{suhr2007fatigue,meza2015resilient} and woven architectures \cite{moestopo2023knots}. 
While the field of architected materials is thriving with advancements on improving specific mechanical properties, the role of the sample geometry of the energy absorbing material and its interplay with the mechanical properties towards meeting critical performance criteria has been overlooked. For example, it is well understood that helmet liners designed to prevent traumatic brain injury during extreme sports or combat require not only high specific energy absorption capacity but also the ability to limit peak accelerations below a critical value \cite{abayazid2024viscoelastic,shuaeib2007new}. 

\begin{figure}[t]
	\centering
	\includegraphics[width=\textwidth]{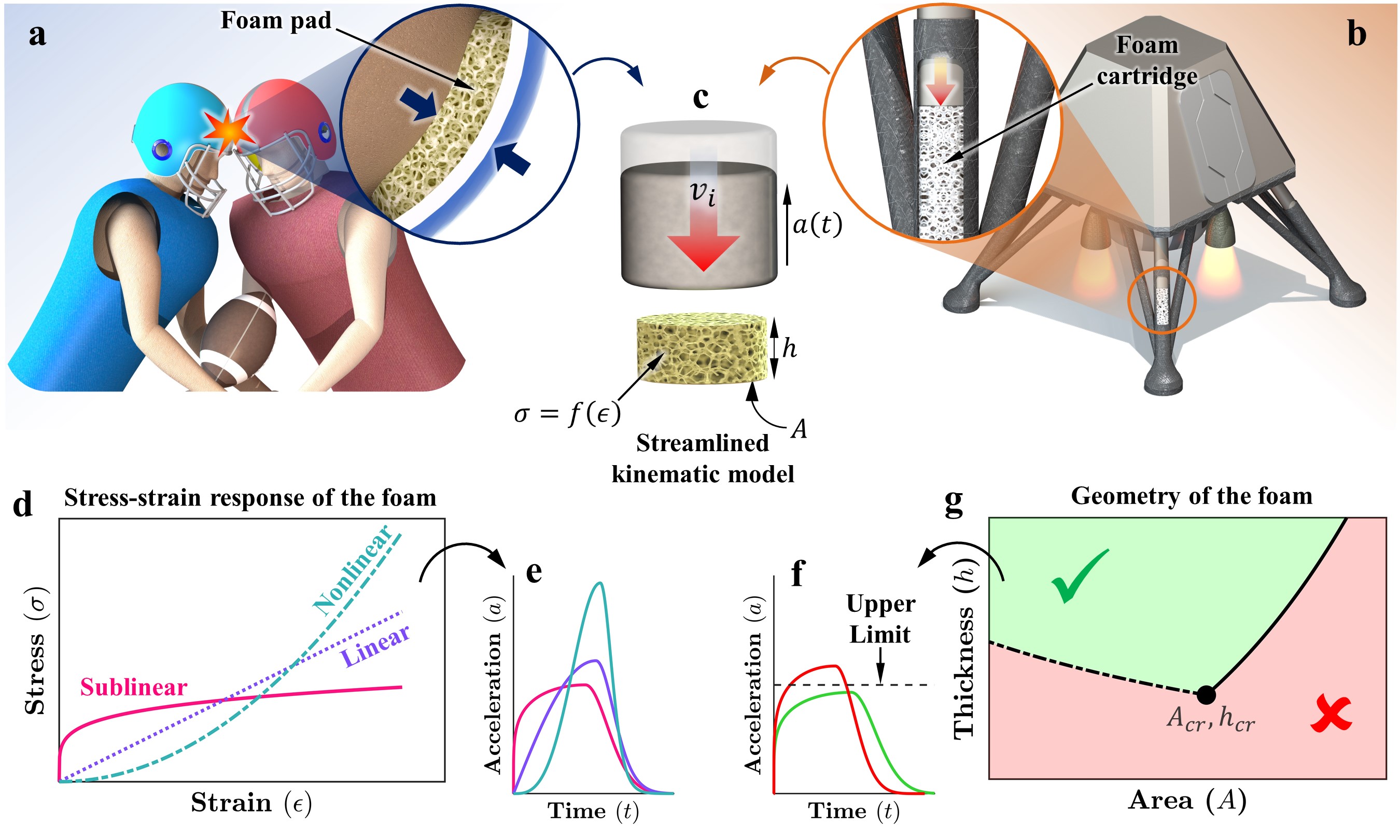}
	\caption{(a) An illustration of a collision in sports, where sudden acceleration can cause Traumatic Brain Injury (TBI). Inset: An energy absorbing pad in a helmet absorbing kinetic energy. (b) An illustration of a spacecraft lander with struts consisting of aluminum honeycomb foam cartridges absorbing residual kinetic energy at touchdown. (c) A streamlined kinematic model showing a mass traveling at a velocity $v_i$ and coming to rest by impacting a foam pad. (d,e) Various types of stress-strain curves seen in cellular materials and their effect on the impact acceleration vs. time curves. (f,g) A design map for the thickness and cross-sectional area of foam to keep the peak acceleration and peak strain below desired limits}
	\label{fig1}
\end{figure}

This interplay between the intrinsic properties of the foam and the protective layer's geometry is governed by three important mechanical properties of foams: relative density ($\bar{\rho}$), the scaling of relative modulus with relative density $(\bar{E}\propto \bar{\rho}^{\alpha})$, and the characteristic shape of the stress-strain response. 
The relative density, which is the ratio of the foam's bulk density ($\rho$) to the density of its solid counterpart ($\rho_s$), controls the compressibility of the foam. A porous foam with a low relative density can be compressed to larger strains before reaching the densification regime beyond which the stress rapidly increases, diminishing the foam's effectiveness.
The relative modulus $(\bar{E})$ of all cellular materials has been observed to scale with relative density $(\bar{E}\propto{\bar{\rho}^\alpha})$. The scaling exponent $(\alpha)$ is determined by the morphology and the deformation mechanism and typically falls within the range $1<\alpha<3$ \cite{chawla2023disrupting,ashby2006properties,zheng2014ultralight}. A linear scaling exponent $(\alpha=1)$ is generally desirable because it allows achieving higher modulus and greater specific energy absorption without significantly increasing the foam's density \cite{chen2019stiff,zheng2014ultralight}. The protective performance of the foams also depends on the characteristic shape of the stress-strain response (\Cref{fig1}(d)). Foams with a plateau-like sublinear stress-strain response are typically considered desirable, allowing absorption of a given amount of energy at a lower stress level. A desirable protective foam, hence, should exhibit high porosity, a near-linear scaling of modulus with density, and a plateau-like sublinear stress-strain curve.

Traditionally, foams exhibiting properties similar to those mentioned above are considered suitable for nearly all protective applications. Contrary to this belief, we report that foams with steeper scaling ($\alpha > 1$) and nonlinear stress-strain behavior can be beneficial in achieving compact and lightweight energy-absorbing pads. Using a streamlined kinematic model and dimensional analysis, we derive guidelines for geometric design and discover optimal mechanical properties resulting in the thinnest and lightest energy absorbing foam pads. Guided by these derived criteria, we design, synthesize, characterize, and demonstrate optimal performance in hierarchically architected vertically aligned carbon nanotube (VACNT) foams.
Freestanding VACNT foams, synthesized via the floating-catalyst thermal chemical vapor deposition (tCVD) process, are renowned for being simultaneously stiff and lightweight, as well as, for exhibiting remarkably high specific energy absorption \cite{cao2005super,thevamaran2015shock}, which rivals that of crashworthy metallic foams and architected foams \cite{chawla2023disrupting}. These properties are attributed to their multi-scale hierarchical structure with the nanoscale interactive fibrous morphology. Unlike metallic foams and architected foams, which often fail in a brittle manner and lack strain recovery, VACNT foams can recover near completely from compressive strains as large as 90\% \cite{cao2005super}. Furthermore, they exhibit high thermal conductivity \cite{chawla2024superior} and maintain consistent mechanical properties across a wide range of strain rates and temperatures \cite{xu2010carbon,raney2013rate}, making them suitable for protective applications in extreme environments.

Using photolithography, we introduced an additional level of structural hierarchy in VACNT foams by creating mesoscale patterns \cite{lattanzi2014geometry,chawla2023disrupting,chawla2022superior}. Incorporating various geometries of mesoscale patterns opens a broad range of tunability in relative density, constitutive response, and density-dependent scaling of mechanical properties. Our hierarchical architecture design of VACNT foams yields optimal pad geometries subjected to imposed constraints on peak acceleration and maximum compressive strain. We demonstrate that while a VACNT foam with a concentric cylinder architecture results in protective pads that limit maximum exerted force by exhibiting a plateau in the stress-strain response, its geometric design space is limited. In contrast, a bio-inspired higher-order fractal architecture and a sparsely packed cylindrical architecture, both exhibiting nonlinear stress-strain responses, enable the design of thin and lightweight protective pads across a much wider range of pad geometries. Our design framework and its effective demonstration on the hierarchically architected VACNT foams provide a comprehensive approach to designing architected materials for superior performance. Our study provides guidance on achieving enhanced protective performance with imposed constraints on geometry and physical properties within the vast material design space that is typically tackled with machine learning \cite{ha2023rapid,liu2022growth} and statistical design of experiments \cite{chawla2022superior}.

\section*{The foundation of the design framework}

Energy-absorbing pads can take various geometries depending on the application, such as a flat comfort pad in helmet liners (\Cref{fig1}(a)) or a long cylindrical cartridge in planetary landers (\Cref{fig1}(b)). In all cases, during an impact, the foam is compressed in the direction of impact, absorbing kinetic energy while minimizing the imparted load and recoil.
The impact scenarios depicted in \Cref{fig1}(a,b) can be described by the simplified equation of motion corresponding to an impacting mass $m$ compressing the protective foam with an initial impact velocity $(v_i)$ resulting in swift acceleration (retardation) to rest (\Cref{fig1}(c)). Assuming that the foam is getting compressed uniformly within an area $A$ and thickness (or height) $h$, the force acting on the mass during acceleration will be equal to the magnitude of force developed in the foam, as follows:

\begin{equation}
    m {\frac{d^2x}{dt^2}}=mg-\sigma_L A,
    \label{eq1}
\end{equation}

with the initial conditions:

\begin{equation}
    { \left.x\right|_{t=0}=0} \;\;,\;\; {\left.\frac{d x}{d t}\right|_{t=0}=v_i} 
    \label{eq1a}
\end{equation}

where $x$ is the compressive displacement in the foam, $t$ represents time, $g$ is the acceleration due to gravity, and $\sigma_L$ is the stress response of the foam, with the subscript $L$ indicating that the foam is being loaded (compressed) by the impact.
We describe the constitutive stress-strain response of the foam by an empirical power law equation \cite{rusch1970load} as follows 

\begin{equation}
    \sigma_L=E_L\epsilon^{\lambda_L}
    \label{eq1b}
\end{equation}

where $E_L$ is elastic modulus, and the exponent $(\lambda_L)$ governs the shape of the constitutive stress-strain response. By varying $\lambda_L$, various stress-strain curves, such as those exhibiting sublinear $(\lambda_L<1)$, linear $(\lambda_L\approx1)$, and nonlinear $(\lambda_L>1)$ behaviors, can be modeled for parametric analysis. Integrating \cref{eq1b} up to a compressive strain $\epsilon_{max}$, we obtain the expression for energy absorption per unit volume $(W_L)$ as follows, 

\begin{equation}
    W_L=E_L {\epsilon_{max}^{\lambda_L+1} \over {\lambda_L +1}} 
    \label{eq1c}
\end{equation}

\begin{figure}[t]
	\centering
	\includegraphics[width=\textwidth]{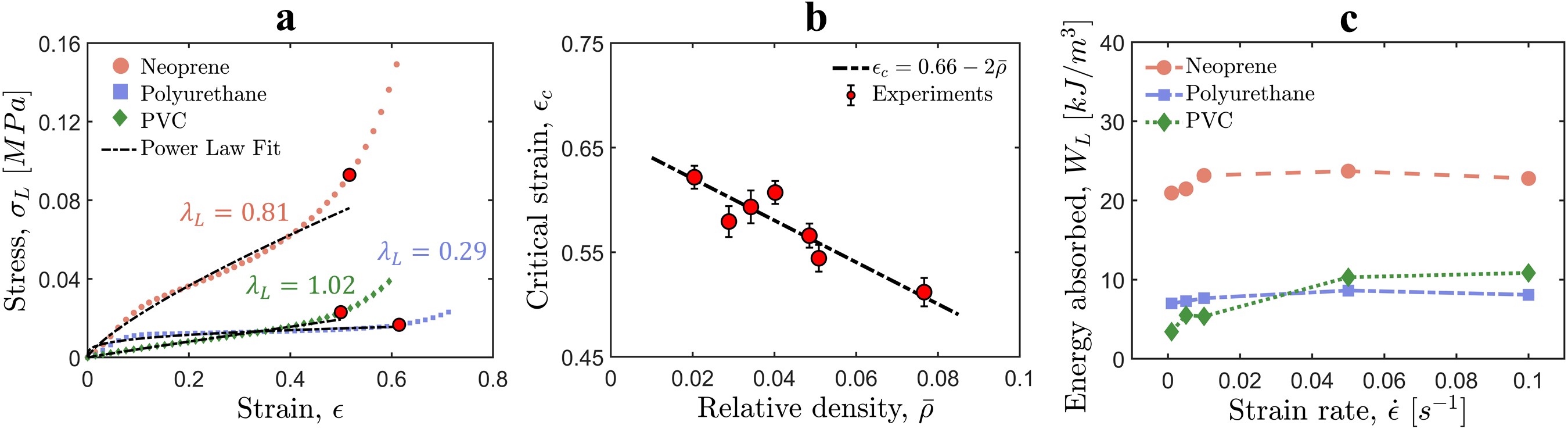}
	\caption{(a) Quasistatic stress-strain curves, measured experimentally at a strain rate of $0.01\;s^{-1}$, for various open-cell polymeric foams, fitted with the power-law model up to the critical strain (indicated by red dots). (b) Critical strain (or onset-to-densification strain) of different polymeric foams plotted as a function of their relative densities. (c) Energy absorbed per unit volume of different foams, measured up to the critical strain, plotted against the strain rate.}
	\label{fig2}
\end{figure}

In the above expression, $W_L$ is inversely proportional to $\lambda_L+1$. This indicates that, for the power-law model, energy absorption significantly decreases as the stress-strain curve becomes more nonlinear (see Figure S1 in \hyperref[section:sd]{SI}). To counteract the effect of $\lambda_L$, we scale the power law by introducing a multiplicative term, $(\lambda_L+1)^{\beta}$, to \cref{eq1b}.

\begin{equation}
    \sigma_L=E_L(\lambda_L+1)^{\beta} \epsilon^{\lambda_L}\;\;,\;\;    W_L=E_L\left(\lambda_L+1\right)^\beta \frac{\epsilon_{max}^{\lambda_L+1}}{\lambda_L+1}
    \label{eq1d}
\end{equation}

where the exponent $\beta\geq1$ determines the relative difference in modulus among a set of foams with various $\lambda_L$ values. The scaled power law (Figure S1 in \hyperref[section:sd]{SI}) more accurately reflects the energy absorption behavior observed in VACNT foams \cite{chawla2023disrupting} and other foam materials \cite{hassani2012production}. When comparing foams with the same $\bar{\rho}$ and $\alpha$, the $\lambda_L$ value primarily governs the shape of the stress-strain response, while its effect on energy absorption is minimal. As shown in \Cref{fig2}(a), the power law equation captures the experimentally measured stress-strain response up to the onset of densification (indicated by red dots)---all measured at the same $0.01\; s^{-1}$ strain rate---fairly well for different open-cell elastomeric foams. These power-law fits yield the effective modulus $(E_L(\lambda_L+1)^{\beta})$ of the foam and a dimensionless exponent $(\lambda_L)$ that describes the shape of the stress-strain curve. The strain at the onset of densification, also called critical strain $(\epsilon_c)$, is where the energy absorption efficiency of the foam reaches the maximum, beyond which it declines rapidly \cite{li2006compressive}. The critical strain $(\epsilon_c)$ is slightly smaller than the actual densification strain $(\epsilon_d)$ where the stress-strain curve becomes almost vertical \cite{gibson2003cellular}. While a foam does keep absorbing more energy for $\epsilon_c<\epsilon<\epsilon_d$, the transmitted force rises sharply making the foam ineffective. Hence, a foam's performance is characterized based on the amount of energy it can absorb before the onset of densification $(\epsilon<\epsilon_c)$. We measured the critical strain of different polymeric foams using the energy absorption efficiency method \cite{li2006compressive} and observe a linear relationship $(\epsilon_c=0.66-2\bar{\rho})$ between the critical strain and the relative density of the foam (\Cref{fig2}(b)). Noteworthy is that this linear fit extrapolates to $\bar{\rho}\to 0.33$ for $\epsilon_c\to 0$, which matches the relative density of foams reported in literature which do not have a plateau region and exhibit densification immediately after the linear elastic regime \cite{gibson2003cellular}. 
In addition to the critical strain, the modulus of the foam also depends on the relative density as mentioned in the introduction:

\begin{equation}
    {{E_L}\over{E_s}}= c_1 \left( {{\rho}\over{\rho_s}} \right)^{\alpha}
    \label{eq2}
\end{equation}

Here, $E_s$ and $\rho_s$ represent the modulus and density, respectively, of the solid material utilized in foam fabrication, $\alpha$ denotes a scaling exponent that is a function of the foam's morphology and deformation mechanism, and $c_1$ is a constant of proportionality \cite{greer2019three}. Utilizing \cref{eq2}, the stress-strain relationship for the foam is expressed as,

\begin{equation}
    \sigma_L= c_1 (\lambda_L+1)^{\beta} E_s \left( {\bar{\rho}} \right)^{\alpha} {\left( x/h\right)}^{\lambda_L},
    \label{eq3}
\end{equation}

where $x/h$ represents the compressive strain $(\epsilon=x/h)$. Similarly, the equation for the energy absorption up to the critical strain $(\epsilon_c)$ is expressed as follows,

\begin{equation}
    {W_L}=c_1 E_s\left(\bar{\rho}^\alpha\right)\left(\lambda_L+1\right)^\beta \frac{\epsilon_{c}^{\lambda_L+1}}{\lambda_L+1}
    \label{eq3_2}
\end{equation}

The stress response in \cref{eq3} is independent of the strain rate, while the most polymeric and elastomeric foams are strain rate sensitive. This rate dependency is generally incorporated into constitutive models by multiplying the elastic term by a viscous damping term \cite{cousins1976theory}. However, we observe that the specific energy absorbed---the area under the stress-strain curve up to the critical strain---as a function of the strain rate we measured experimentally on various open-cell elastomeric foams show only a mild strain rate effect with it initially increasing and then becoming almost constant (\Cref{fig2}(c)). 

In \Cref{fig3b}(a), we compare the stress-strain response of open-cell polyurethane foam for a quasistatic strain rate of $0.1\;s^{-1}$ and dynamic strain rates of $2000\;s^{-1}$ and $2800\;s^{-1}$. From $0.1\;s^{-1}$ to $2000\;s^{-1}$, the overall plateau stress level increases by $1.6\times$. However, for a further increase in strain rate, the response shows negligible change, reaching a steady-state response with minimal additional rate effects \cite{bhagavathula2022density,ouellet2006compressive}. Given the mild dependence of the stress-strain response on the strain rate, the rate-independent power-law (\Cref{eq3}) can still be used to model the stress-strain response that is stable within a working range of strain rate for an application. Moreover, our VACNT foams exhibit rate-independent behavior from quasistatic to very large strain rates \cite{xu2010carbon,raney2013rate}. Therefore, adding a rate-dependent term in \cref{eq3} is not worth considering at the cost of simplicity. 
To make our modeling scale-free, we establish the following dimensionless variables:

\begin{equation}
    \bar{x}=x\times {{a_{c}}\over{v_i^2}} \;\;,\;\;\bar{t}=t\times {{a_{c}}\over{v_i}}  \;\;,\;\; \bar{h} = h\times {{2 a_{c}}\over{v_i^2}} \;\;,\;\; \bar{A} = A\times {{E_s}\over{ma_{c}}}
    \label{eq4}
\end{equation}

Here, $a_{c}$ represents the maximum acceleration limit that the protected object should not exceed. This limit depends on the object's structural resilience or physiological tolerance, for example, the allowable peak acceleration in the case of traumatic brain injury prevention \cite{carlsen2021quantitative}. By substituting the dimensionless variables (\Cref{eq4}) and stress-strain relation (\Cref{eq3}) into \cref{eq1} and \cref{eq1a}, we arrive at the following dimensionless governing equation and dimensionless initial conditions:

\begin{equation}
    {{d^2\bar{x}}\over{d\bar{t}^2}} - {{g}\over{a_{c}}}=-c_1 \bar{A} (\lambda_L+1)^{\beta} \left( {\bar{\rho}} \right)^{\alpha} {\left( 2\bar{x}/\bar{h}\right)}^{\lambda_L}
    \label{eq6}
\end{equation}

\begin{equation}
    { \left.\bar{x}\right|_{t=0}=0} \;\;,\;\; {\left.\frac{d \bar{x}}{d \bar{t}}\right|_{t=0}=1} 
    \label{eq7}
\end{equation}

Usually $a_{c}\gg g$, so the term $(g/a_{c})$ can be ignored \cite{mustin1968theory}. When the impacting mass accelerates to rest from initial velocity $v_i$ as the foam absorbs all its kinetic energy, the maximum compression $(\bar{x}_{max})$ experienced by the foam can be calculated as follows (see \hyperref[section:sd]{SI} for details):

\begin{equation}
    {\left( 2 {{\bar{x}_{max}}\over{\bar{h}}}\right)}^{\lambda_L+1} = {{\lambda_L+1}\over c_1{\bar{A}\bar{h} (\lambda_L+1)^{\beta} (\bar{\rho})^{\alpha}}}
    \label{eq8}
\end{equation}

The magnitude of acceleration will also reach its peak when the compressive strain in the foam reaches its maximum. From \cref{eq6}, the expression for maximum acceleration can be obtained as,

\begin{equation}
    {\left|{d^2\bar{x}}\over{d\bar{t}^2}\right|_{max}}=c_1 \bar{A} (\lambda_L+1)^{\beta} \left( {\bar{\rho}} \right)^{\alpha} {\left( {{2}\over {\bar{h}}} \bar{x}_{max}\right) }^{\lambda_L}
    \label{eq9}
\end{equation}

By substituting $\bar{x}_{max}$ from \cref{eq8}, we obtain the following expression for the magnitude of peak acceleration:

\begin{equation}
    {\left|{d^2\bar{x}}\over{d\bar{t}^2}\right|_{max}}= \left( c_1 \bar{A} (\lambda_L+1)^{\beta} \left( {\bar{\rho}} \right)^{\alpha} \right)^{1/(\lambda_L+1)} \times {\left({\lambda_L+1}\over{\bar{h}}\right) }^{\lambda_L/{\lambda_L+1}}
    \label{eq9b}
\end{equation}

Our objective is to minimize both the mass and the thickness of the foam required to absorb the kinetic energy of the impact while ensuring that the magnitude of the maximum acceleration $(a_{max})$ stays below the desired limit $(a_c)$ and the maximum compressive strain $(\epsilon_{max})$ remains below the critical strain $(\epsilon_c)$.

\begin{equation}
   \epsilon_{max}={x_{max} \over h} \leq \epsilon_c \;\;\; \rightarrow  \;\;\; {{\bar{x}_{max}}\over \bar{h}}\leq {\epsilon_c \over 2}
   \label{eq10}
\end{equation}

\begin{equation}
    a_{max} = \left|\frac{d^2 x}{d t^2}\right|_{\max } \leq a_{c} \;\;\;\rightarrow \;\;\; \left|\frac{d^2 \bar{x}}{d \bar{t}^2}\right|_{\max } \leq 1
    \label{eq11}
\end{equation}

By substituting \cref{eq8} and \cref{eq9b} into \cref{eq10} and \cref{eq11}, respectively, we obtain the following constraints on $\bar{h}$ for a given value of dimensionless foam area $\bar{A}$ and parameters describe the foam's mechanical properties $(c_1,\lambda_L,\alpha,\bar{\rho},\epsilon_c,\beta)$

\begin{equation}
{\bar{h}} \geq {{(\lambda_L+1)}^{1-\beta} \over {c_1 \bar{A} {{(\bar{\rho})}^{\alpha}} {{(\epsilon_c)}^{\lambda_L+1}} }}  
\label{eq13} \;\;\;\; (\epsilon_{max}\leq \epsilon_c)
\end{equation}

\begin{equation}
\bar{h} \geq (\lambda_L+1)^{\left( 1+{\beta \over \lambda_L} \right)} (c_1 \bar{A})^{1/\lambda_L} (\bar{\rho})^{\alpha/\lambda_L} \;\;\;\; (a_{max}\leq a_c)
\label{eq14}
\end{equation}

\Cref{eq13} and \cref{eq14} establish individual conditions on $\bar{h}$ to satisfy the criteria for maximum strain and acceleration, respectively. However, both equations are also functions of $\bar{A}$. Therefore, to derive a condition on $\bar{h}$ that depends solely on material properties, we solve the two inequalities to obtain the individual constraints on both $\bar{h}$ and $\bar{A}$ as follows (refer to \hyperref[section:sd]{SI} for detailed derivation):

\begin{equation}
    \bar{h} \geq  \bar{h}_{cr}\;\;,\;\; \bar{h}_{cr}={{\lambda_L+1}\over{\epsilon_c}}
    \label{eq15}
\end{equation}

\begin{equation}
    \left( \bar{h} \over \bar{h}_{cr} \right) ^{\lambda_L} \times \bar{A}_{cr} \geq \bar{A} \geq \left( \bar{h}_{cr} \over \bar{h} \right)\times \bar{A}_{cr} \;\;,\;\; \bar{A}_{cr} = {1\over {c_1 (\lambda_L+1)^{\beta} {\epsilon_c}^{\lambda_L} {\bar{\rho}^{\alpha}}}}
    \label{eq16}
\end{equation}

\begin{figure}[t]
	\centering
	\includegraphics[width=\textwidth]{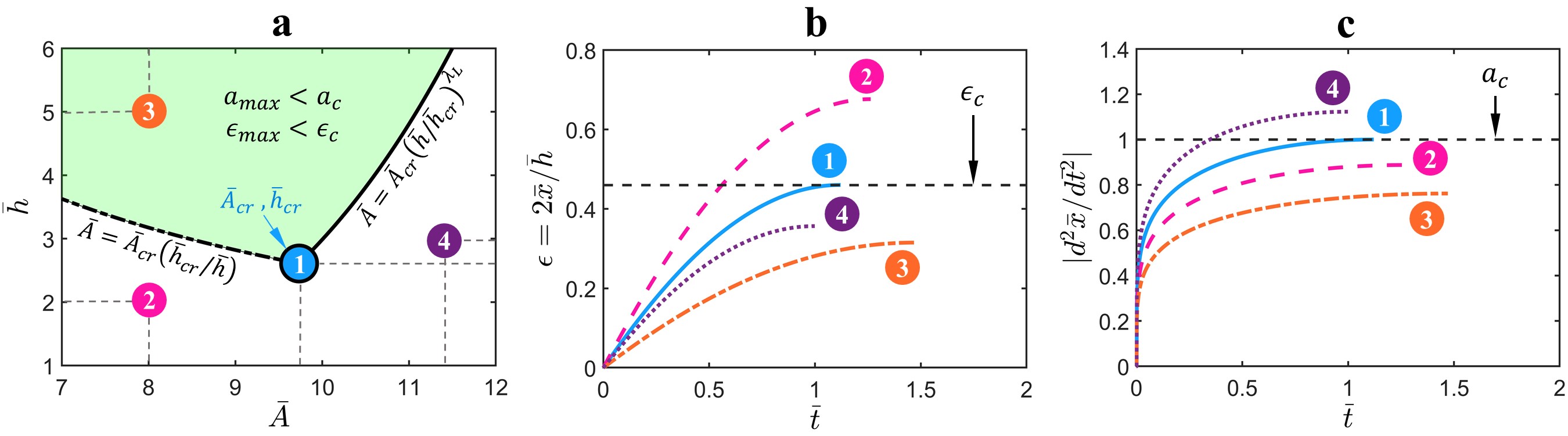}
	\caption{(a) Geometric design space showing constraints on dimensionless thickness $(\bar{h})$ and dimensionless area $(\bar{A})$ ($\alpha=1$, $\lambda_L=0.2$, $\beta=1$, $\bar{\rho}=0.1$, $c_1=1$, $\epsilon_c=0.66-2\bar{\rho}$). In the shaded region, both peak acceleration and peak strain will remain below the desired limits for any combination of $\bar{A}$ and $\bar{h}$, while absorbing the given kinetic energy of impact. (b) Compressive strain in the foam during impact as a function of dimensionless time $(\bar{t})$. The maximum limit on strain (critical strain, $\epsilon_c$) is indicated by a black dashed line. (c) Dimensionless acceleration as a function of dimensionless time. The limit on maximum acceleration is shown by a black dashed line ($a_c$).}
	\label{fig3}
\end{figure}

\begin{figure}[t]
	\centering
	\includegraphics[width=\textwidth]{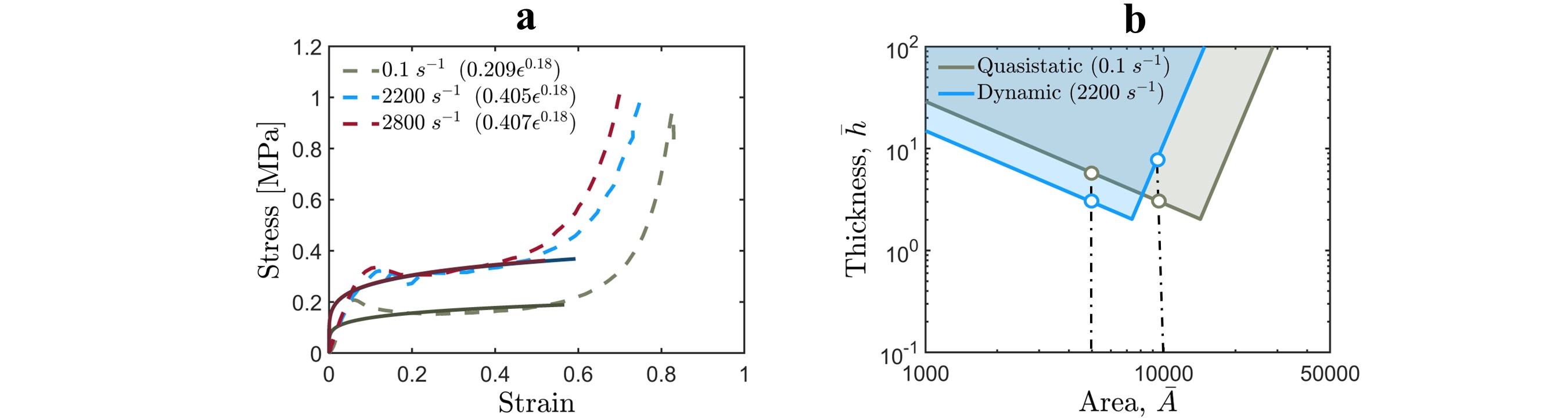}
	\caption{(a) Effect of strain rate on the experimentally measured stress-strain response of an open-cell polyurethane foam, along with the corresponding power-law fits. (b) Geometric design space for quasistatic and dynamic strain rates, obtained using power-law fits with the assumption $E_s = 2.7\; \textnormal{GPa}$ for solid polyurethane.}
	\label{fig3b}
\end{figure}

The above inequalities suggest that, to maintain the peak acceleration and maximum compression strain within the desired limits (\Cref{eq13}), the foam's thickness must be greater than a critical value $\bar{h}_{cr}$, which depends on the material properties, regardless of $\bar{A}$. Conversely, the cross-sectional area can vary within a range determined by the ratio $\bar{h}/\bar{h}_{cr}$ and a critical area $\bar{A}_{cr}$. For the limiting case of $\bar{h}=\bar{h}_{cr}$, $\bar{A}$ must exactly be equal to $\bar{A}_{cr}$. The flexibility in selecting $\bar{A}$ and $\bar{h}$ provides design freedom for addressing different challenging applications with geometric constraints on shock absorbers. In \Cref{fig3}(a), we illustrate such combinations in a shaded region bounded by the upper and lower limits of $\bar{A}$ for an example $\bar{A}_{cr}$ and $\bar{h}_{cr}$ calculated from a certain set of material parameters ($\alpha=1$, $\lambda_L=0.2$, $\beta=1$, $\bar{\rho}=0.1$, $c_1=1$, $\epsilon_c=0.66-2\bar{\rho}$). All combinations of $\bar{A}$ and $\bar{h}$ that fall within the green-shaded region will ensure that the peak acceleration and peak compressive strain remain below the desired limits while the entire kinetic energy due to impact is absorbed by the foam. To validate this, we select four different combinations of $\bar{A}$ and $\bar{h}$, as marked in \Cref{fig3}(a), for which we solve the time-domain differential equation (\Cref{eq6}). In \Cref{fig3}(b) and \Cref{fig3}(c), we plot strain and dimensionless acceleration, respectively, as functions of dimensionless time. At the critical point $(\bar{A}=\bar{A}_{cr}, \bar{h}=\bar{h}_{cr})$, the peak acceleration and the maximum strain exactly match the set upper limits $a_c$ and $\epsilon_c$. For points (2) and (4), one of the conditions is not satisfied, whereas for point (3), which lies in the shaded region, both conditions are satisfied. We further validated our geometric design criteria by conducting impact simulations on a metallic 3D-architected foam using Abaqus CAE. Simulations were performed for various pad geometries to compare peak acceleration and peak compressive strain (see Figure S2 in \hyperref[section:sd]{SI}). Our results confirmed that for pad geometries with dimensions lying within the shaded region of the geometric design space, both peak acceleration and peak compressive strain remained within the specified limits.

Considering that the geometric design space shown in \Cref{fig3}(a) is plotted for dimensionless variables $\bar{h}$ and $\bar{A}$, changes in the parameters used to nondimensionalize these variables $(v_i, a_c, m)$ will result in the geometric design space being shifted and scaled along the x and y axes. Similarly, the material's strain rate dependence can also alter the geometric design space. For example, when fitting \cref{eq3} to the stress-strain response of polyurethane foam (\Cref{fig3b}(a)), we observed that as the strain rate increases, the modulus increases $(E_L)$ while keeping $\lambda_L$ and $\epsilon_c$ nearly constant. Consequently, the increase in strain rate only proportionally reduces $\bar{A}_{cr}$, while $\bar{h}_{cr}$ remains constant (\Cref{eq15}, \Cref{eq16}). This results in the geometric design space shifting leftward, affecting the minimum allowable $\bar{h}$ for a given $\bar{A}$, as shown with two dotted lines in \Cref{fig3b}(b). For a foam material with modulus characterized as a function of strain rate, all the geometric design space for various strain rate scenarios can be obtained (also see Figure S9 in \hyperref[section:sd]{SI}).

\section*{Thickness and mass minimization}

As shown in the previous section, for a given set of material parameters $(\alpha, \lambda_L, \beta, \bar{\rho}, c_1)$, the thickness of the foam $(\bar{h})$ must be greater than the critical thickness $(\bar{h}_{cr})$, while the cross-sectional area $(\bar{A})$ can fall within a broad range defined by a lower and an upper limit (\Cref{eq16}) in order to limit peak acceleration and maximum compressive strain while entirely absorbing the kinetic energy of impact. These bounds on cross-sectional area depend on the ratio $\bar{h}/\bar{h}_{cr}$, such that the higher the ratio, the broader the range. Thus, any desired cross-sectional area $(\bar{A})$ within the permissible range can be achieved by proportionally scaling the thickness. We can use this property to identify an optimal set of material parameters that will minimize the thickness of a protective foam given a specific cross-sectional area, for absorbing a given kinetic energy of impact.

\begin{figure}[ht!]
	\centering
	\includegraphics[width=\textwidth]{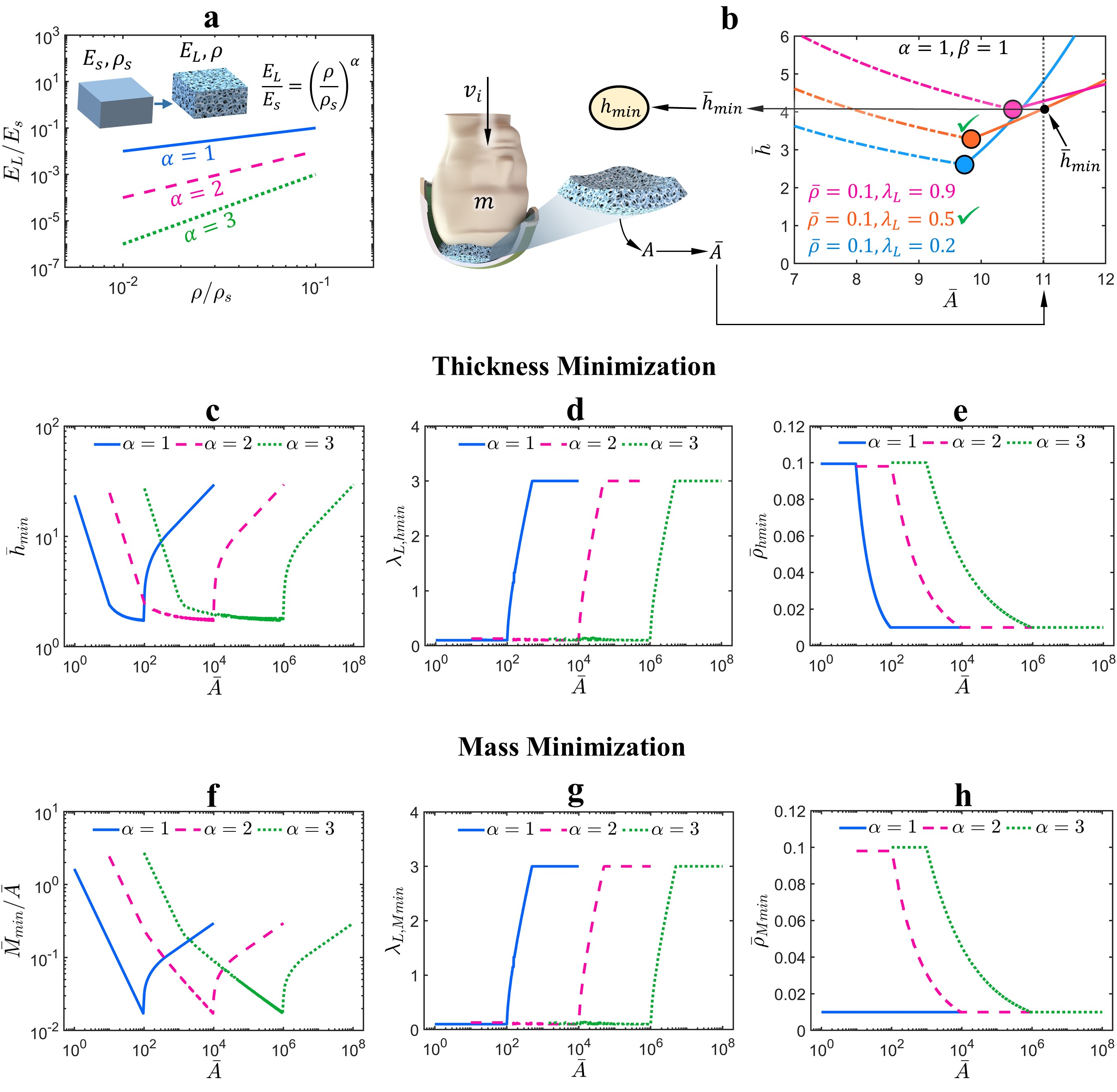}
	\caption{(a) Scaling of the relative modulus with the relative density. (b) Determining the minimum possible thickness for a protective foam pad of a known area $\bar{A}$ from a given set of foam material parameters; inset shows a surrogate head-helmet system undergoing impact. (c) Minimum thickness as a function of area for material parameters $0.01<\bar{\rho}<0.1$, $0.1<\lambda_L<3$, and $1<\alpha<3$. (d,e) Values of $\lambda_L$ and $\bar{\rho}$ corresponding to the minimum thickness. (f) Minimum mass per unit area as a function of area for material parameters $0.01<\bar{\rho}<0.1$, $0.1<\lambda_L<3$, and $1<\alpha<3$. (g,h) Values of $\lambda_L$ and $\bar{\rho}$ corresponding to the minimum mass per unit area.}
	\label{fig4}
\end{figure}

To this end, we consider a foam with density $\rho$ and modulus $E_L$ made of a solid material with density $\rho_s$ and modulus $E_s$. As the relative modulus scales with relative density (\Cref{fig4}(a)), we explore three different scaling exponents $(\alpha)$ that are commonly found in cellular materials literature: a linear scaling $(\alpha=1)$ that is typically associated with a stretch-dominated deformation mechanism of the foam's micro-structure \cite{ashby2006properties}, a quadratic scaling $(\alpha=2)$ observed in bending-dominated deformation mechanisms \cite{schaedler2011ultralight}, and cubic scaling $(\alpha=3)$ which has been observed in foams with stochastic micro-structure \cite{worsley2009mechanically}. Values of $\alpha$ greater than 3 are exceedingly rare \cite{gross1997elastic}. While certain efficient architectures demonstrate $\alpha\approx1$ \cite{chen2019stiff,novak2021quasi}, values of $\alpha$ less than $1$ have not been reported. It is worth noting that many stretch-dominated lattices exhibit a yield transition after the initial elastic regime, characterized by a peak stress and subsequent softening before reaching a plateau of nearly constant stress. While our empirical two-parameter monotonic power-law function may not fully capture this behavior, it still approximates the overall sublinear trend and provides a reasonable prediction of energy absorption and acceleration (see Figure S3 in \hyperref[section:sd]{SI}). Additionally, some lattices show linear scaling with either no peak stress or negligible peak stress \cite{zheng2014ultralight,chen2019stiff,chawla2023disrupting}, thus our choice to begin the range of $\alpha$ from $\alpha = 1$ is motivated by the commonly observed scaling behavior in cellular materials.

We assume that the relative density varies between $\bar{\rho}=0.01$ and $\bar{\rho}=0.1$, which is typical for open-cell foams \cite{gibson2003cellular}. Also this is approximately the range of $\bar{\rho}$ for which we measured the critical strain $(\epsilon_c=0.66-2\bar{\rho})$ of various open-cell polymeric foams (\Cref{fig2}(b)). The measured critical strain $(\epsilon_c)$ follows a linear relationship with relative density, as shown in \Cref{fig2}(b). Additionally, we assume that $\lambda_L$ ranges from $\lambda_L=0.1$ (indicative of a highly sublinear stress-strain response resembling a Heaviside function) to $\lambda_L=3$ (indicative of a highly nonlinear stress-strain response) (\Cref{fig1}(d)). 

The workflow to achieve minimum thickness for a given cross-sectional area is illustrated in \Cref{fig4}(b) using an example of a head-helmet system undergoing a blunt impact, similar to that of a drop tower test for evaluating helmet performance. Here, $m$ represents the effective mass of the head-helmet system, while $v_i$ denotes the impact velocity.
While the foam liner---often in combination with an additional soft layer for comfort---inside the helmet undergoes much more complex deformation \cite{maheswaran2024mitigating}, the direct impact shown in \Cref{fig4}(b) leads to axial compression of the foam liner that absorbs kinetic energy. For the cross-sectional area $A$ of the pad that experiences direct impact, the corresponding dimensionless area can be calculated using \cref{eq4}. The \Cref{fig4}(b) illustrates the process of obtaining the minimum thickness $(\bar{h}_{min})$ for a given $\bar{A}$ ($\bar{A}=11$ in this example), indicated by a black dotted vertical line. 
Here, we consider three different $\bar{h}_{cr}$ values and their corresponding $\bar{A}_{cr}$ values (represented as colored dots in \Cref{fig4}(b)), which were obtained for $\bar{\rho}$, $\lambda_L$, and $\alpha$ within the parameter space we considered ($0.01<\bar{\rho}<0.1$, $0.1<\lambda_L<3$, $\alpha=1,2,3$). 
For each dot, when $\bar{h}$ is scaled, the lower and upper limits of permissible $\bar{A}$ are depicted by dashed and solid lines, respectively. The solid lines extend to our target $\bar{A}=11$, intersecting the vertical dotted line at different locations. Among the three dots, the orange dot corresponding to $\bar{\rho}=0.1$ and $\lambda_L=0.5$ results in the minimum thickness $(\bar{h}_{min})$. 
This dimensionless minimum thickness can be converted to dimensional minimum thickness $(h_{min})$ using the known parameters $v_i$ and $a_c$ to fabricate a compact foam pad (\Cref{eq4}). Let's denote the $\bar{\rho}$ and $\lambda_L$ corresponding to the minimum thickness as $\bar{\rho}_{hmin}$ and $\lambda_{L,hmin}$, respectively. The chosen energy absorbing foam must have a relative density equal to $\bar{\rho}_{hmin}$ and a stress-strain response with $\lambda_L= \lambda_{L,hmin}$ to achieve the minimum thickness.

\begin{figure}[t]
	\centering
	\includegraphics[width=\textwidth]{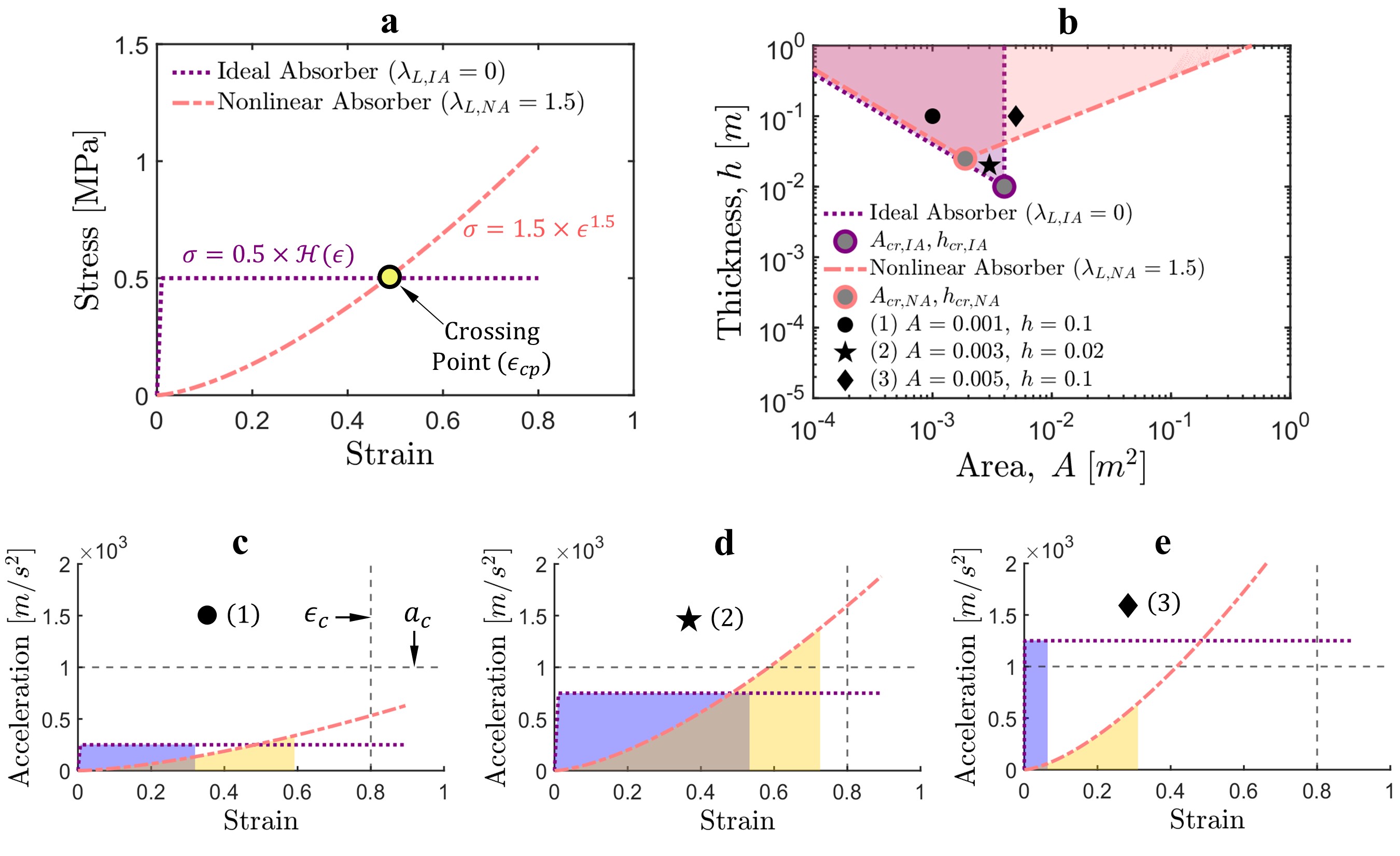}
	\caption{(a) Stress strain response of an ideal absorber vs. a nonlinear absorber. (b) Geometric design space of ideal absorber vs. nonlinear absorber plotted assuming $\epsilon_c=0.8,\;E_s=10 MPa$, $m_s=2\;kg,\;v_i=4\;m/s,\;E_{L,NA}(\lambda_{L,NA}+1)^{\beta}=1.48\;MPa,\;a_c=1000\;m/s^2$. (c,d,e) Acceleration vs strain for various pad geometries.}
	\label{fig5b}
\end{figure}

Similarly, all possible pairs of $\bar{h}_{cr}$ and $\bar{A}_{cr}$ within the range of material parameters ($0.01<\bar{\rho}<0.1$, $0.1<\lambda_L<3$, $\alpha=1,2,3$) can be compared for a given $\bar{A}$ to determine the absolute minimum thickness.
In \Cref{fig4}(c), we plot the absolute minimum thickness as a function of $\bar{A}$ obtained for material parameters $\bar{\rho}$, $\lambda_L$, and $\alpha$, varying within the range we assumed, while $\beta=1$ is fixed. As $\bar{A}$ increases, $\bar{h}_{min}$ counters $\bar{A}$ and decreases initially to maintain the volume of the foam nearly constant for energy absorption, reaching a minimum, and then sharply rises to prevent the effective stiffness $(E_LA/h)$ of the foam pad from becoming too large. The descending portion is governed by the constraint on $\bar{h}$ set in \cref{eq13}, whereas the ascending portion is governed by \cref{eq14} (see Figure S4(c) in \hyperref[section:sd]{SI}). Notably, the condition $\bar{h}\geq\bar{h}_{cr}$ is satisfied in both the ascending and descending portions. In the middle section, the minimum thickness becomes equal to the critical thickness, i.e., $\bar{h}_{min} = \bar{h}_{cr}$. While the equation for $\bar{h}_{cr}$ is independent of $\bar{A}$, $\lambda_{L,hmin}$ and $\bar{\rho}_{hmin}$ vary as functions of $\bar{A}$, resulting in a curved middle section in \Cref{fig4}(c). For smaller cross-sectional areas $(\bar{A}<100)$, linear scaling $(\alpha=1)$ results in the smallest thickness, whereas for $\bar{A}>10^4$, cubic scaling is better. This suggests that linear scaling results in a compact-sized foam pad for applications with space limitations in terms of area, such as a foam cartridge in landing struts (\Cref{fig1}(b)). On the other hand, cubic scaling performs better when more area has to be covered with foam, such as helmet liners and packaging applications.

\Cref{fig4}(d,e) show optimized parameters corresponding to the minimum thickness $\bar{h}_{min}$. A small value of $\lambda_{L}$ initially leads to a lower value of $\bar{h}_{cr}$ (\Cref{eq15}). Therefore, it's not surprising that for a smaller cross-sectional area, the thickness is minimized for $\lambda_{L,hmin}=0.1$. In contrast, for higher $\bar{A}$, the minimum thickness occurs for $\lambda_{L,hmin}=3$. Notice the power exponent $\lambda_L$ in the upper limit of $\bar{A}$ in \cref{eq16}, which allows for attaining significantly higher $\bar{A}$ without scaling $\bar{h}$ by a large amount. This demonstrates the utility of foams with a nonlinear stress-strain curve $(\lambda_L>1)$, which contradicts the conventionally held belief that foams should always exhibit a sublinear response with a plateau of nearly constant stress. 

To elucidate this phenomenon, we compare the geometric design space of an ideal shock absorber---a foam with a stress-strain response that resembles a Heaviside step function ($\lambda_{L,IA}=0$) to a nonlinear absorber $(\lambda_{L,NA}=1.5)$ (\Cref{fig5b}). Generally, it is argued that the presence of a plateau makes a foam superior in shock absorption because it limits the amount of transmitted force for large compressive strains. However, such a feature constrains the design space for pad's geometry. \Cref{fig5b}(a) shows the stress-strain responses of an ideal absorber with a sharp rise in stress followed by a plateau of constant stress and a nonlinear absorber that exhibits a slow rise in stress eventually crossing the ideal absorber's stress-strain curve at $\epsilon_{cp}$. \Cref{fig5b}(b) compares the geometric design space of the ideal absorber with that of the nonlinear absorber. While the ideal absorber provides thinner shock-absorbing pads for smaller cross-sectional areas $({A}<{A}_{cr,IA})$, its design space abruptly terminates $({A}={A}_{cr,IA})$ for larger areas $({A}>{A}_{cr,IA})$. In contrast, the nonlinear absorber allows the creation of thin and lightweight energy-absorbing pads over a broader range of cross-sectional areas $({A}>{A}_{cr,IA})$. 

Although the plateau in the ideal absorber limits the maximum possible stress, the initial sharp rise prevents the stress from remaining low, even at small strains. To illustrate this, we selected three combinations of $A$ and $h$ in \Cref{fig5b}(b), representing three distinct pad geometries. \Cref{fig5b}(c,d,e) shows the acceleration as a function of strain for these geometries. The shaded regions beneath the curves indicate the extent of foam compression, for the given kinetic energy of impact. For pad geometry (1), the maximum compression strain and peak acceleration remain well below the critical limits for both the ideal and nonlinear absorbers. For pad geometry (2), because the thickness lies outside the geometric design space of the nonlinear absorber, the acceleration exceeds the critical limit. In contrast, the ideal absorber maintains a lower acceleration due to its plateau. For pad geometry (3), where $A > A_{cr,IA}$, the ideal absorber becomes excessively stiff, causing the acceleration to exceed the upper limit. However, the nonlinear absorber, with its initial gradual increase in stress, absorbs the same amount of energy while keeping the acceleration below the upper limit. Due to this, it offers greater versatility in geometric design, enabling the thickness to be adjusted so that the maximum strain remains below $\epsilon_{cp}$ (\Cref{fig5b}(a)). It is noteworthy that the ideal absorber, while exhibiting higher specific energy absorption for strains below $\epsilon_{cp}$, it becomes too stiff to be used for $A>A_{cr,IA}$. The importance of this fundamental understanding is best exemplified by recent work on graded architected foams \cite{yu2019investigation} and the development of bi-layer (soft + hard) combat helmet liners \cite{koumlis2019strain}, which aim to achieve a gradual initial rise in stress similar to our nonlinear absorber.

In \Cref{fig4}(e), the relative density corresponding to $\bar{h}_{min}$ $(\bar{\rho}_{hmin})$ on the other hand follows an opposite trend to $\lambda_{L,hmin}$. A higher relative density for a small $\bar{A}$ provides stiffness to the foam pad, whereas a low relative density, which results in a large critical strain $(\epsilon_c=0.66-2\bar{\rho})$, limits $\bar{h}_{cr}$ (\cref{eq15}) for large $\bar{A}$. It is worth mentioning that $\bar{h}_{min}$ and $\bar{A}$ are dimensionless; therefore, their magnitudes in \Cref{fig4}(c) are not to scale. Their magnitudes are meaningful only when converted back to their dimensional form using \cref{eq4}. Moreover, since we made our equations scale-free by rendering them dimensionless beforehand, any variation in the external factors such as $v_i$, $m$, and $a_c$ will only scale up or scale down $\bar{A}$ and $\bar{h}$ in \Cref{fig4}(c) without changing the trend of the curves. For example, setting a lower value of $a_c$, which is akin to increasing the factor of safety, will make $\bar{A}$ larger for a constant $A$ (\Cref{eq4}), affecting only the selected $\bar{h}_{min}$ (\Cref{fig4}(c)) along with the associated $\lambda_{L,hmin}$ and $\bar{\rho}_{hmin}$.

\begin{figure}[t]
	\centering
	\includegraphics[width=\textwidth]{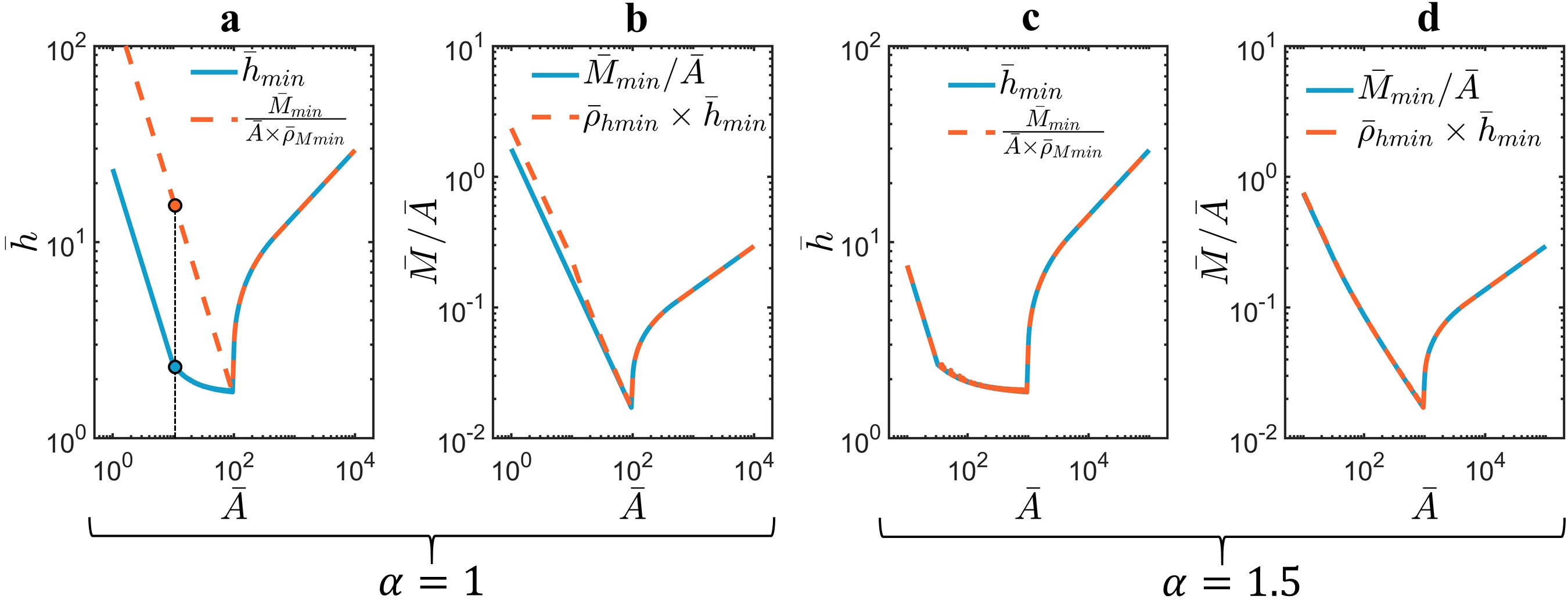}
	\caption{(a) Comparison of the absolute minimum thickness with the thickness associated with the minimum mass for $\alpha=1$. (b) Comparison of the absolute minimum mass per unit area with the mass per unit area associated with the minimum thickness for $\alpha=1$. (c) Comparison of the absolute minimum thickness with the thickness associated with the minimum mass for $\alpha=1.5$. (d) Comparison of the absolute minimum mass per unit area with the mass per unit area associated with the minimum thickness for $\alpha=1.5$.}
	\label{fig5}
\end{figure}

Minimizing thickness and making a foam pad compact in size doesn't always guarantee minimum possible mass. The mass depends on both the volume and the relative density. The dimensionless mass can be calculated using the following equation 

\begin{equation}
    \bar{M} = \bar{\rho}\bar{A}\bar{h}
\end{equation}

The above mass can be minimized and the resultant optimal material parameters can be obtained by using the similar methodology we implemented earlier for $\bar{h}_{min}$ (\Cref{fig4}(b)). Since mass scales with area, we have plotted the minimum mass per unit area $(\bar{M}_{min}/\bar{A})$ in \Cref{fig4}(f). The trend is similar to $\bar{h}_{min}$: linear scaling $(\alpha=1)$ performs better for a small cross-section area, whereas cubic scaling is better when overlaying a larger area with foam. In \Cref{fig4}(g,h), we plot the optimized parameters $\lambda_{L,Mmin}$ and $\bar{\rho}_{Mmin}$ corresponding to minimum mass. For $\alpha=2$ and $\alpha=3$, both $\lambda_{L,Mmin}$ and $\bar{\rho}_{Mmin}$ exactly match $\lambda_{L,hmin}$ and $\bar{\rho}_{hmin}$ respectively which suggest that both mass and thickness are simultaneously minimized. However, for $\alpha=1$, while $\lambda_L$ still matches, the relative densities are different. $\bar{\rho}_{Mmin}$ remains constant at $0.01$, which is the smallest relative density we considered. This occurs because for $\alpha=1$, the ${\bar{\rho}}^{\alpha-1}$ term vanishes from the critical mass, which is defined as follows:

\begin{equation}
    \bar{M}_{cr} = \bar{\rho}\bar{A}_{cr}\bar{h}_{cr} =  {1\over {c_1 (\lambda_L+1)^{\beta-1} {\epsilon_c}^{\lambda_L+1} {\bar{\rho}^{\alpha-1}}}}
\end{equation}

While $\epsilon_c$ is a function of relative density, it is maximum for $\bar{\rho}=0.01$ (\Cref{fig2}(b)), thus minimizes $\bar{M}_{cr}$. 

To summarize, a nonlinear scaling of relative modulus with relative density allows for the simultaneous minimization of both the mass and thickness of the foam. In contrast, for linear scaling $(\alpha=1)$, minimizing one disregards the other, not allowing both objectives to be simultaneously achieved. We can witness this by comparing the absolute minimum thickness $(\bar{h}_{min})$ with the thickness corresponding to the minimum mass, expressed as follows:

\begin{equation}
   \bar{h}_{min}  \;\;\; \textrm{vs.} \;\;\; \bar{h} = { \bar{M}_{min} \over {\bar{A}\times \bar{\rho}_{Mmin}}}
\end{equation}

\Cref{fig5}(a) illustrates that for $\alpha=1$, minimizing the mass has unintentionally resulted in a much higher thickness. For example, at $\bar{A}=10$ (black dotted line), the absolute minimum thickness is $\bar{h}_{{min}}\approx 2.38$ (on blue solid curve), while the thickness corresponding to when mass is minimized (ignoring thickness) is ${ { \bar{M}_{min} \over {\bar{A}\times \bar{\rho}_{Mmin}}}}\approx16.35$ (orange dashed curve).
Similarly, minimizing the thickness leads to a slightly higher mass, as shown in \Cref{fig5}(b). For $\alpha\geq1.5$, both mass and thickness are simultaneously minimized (\Cref{fig5}(c,d)). 

\begin{figure}[ht!]
	\centering
	\includegraphics[width=\textwidth]{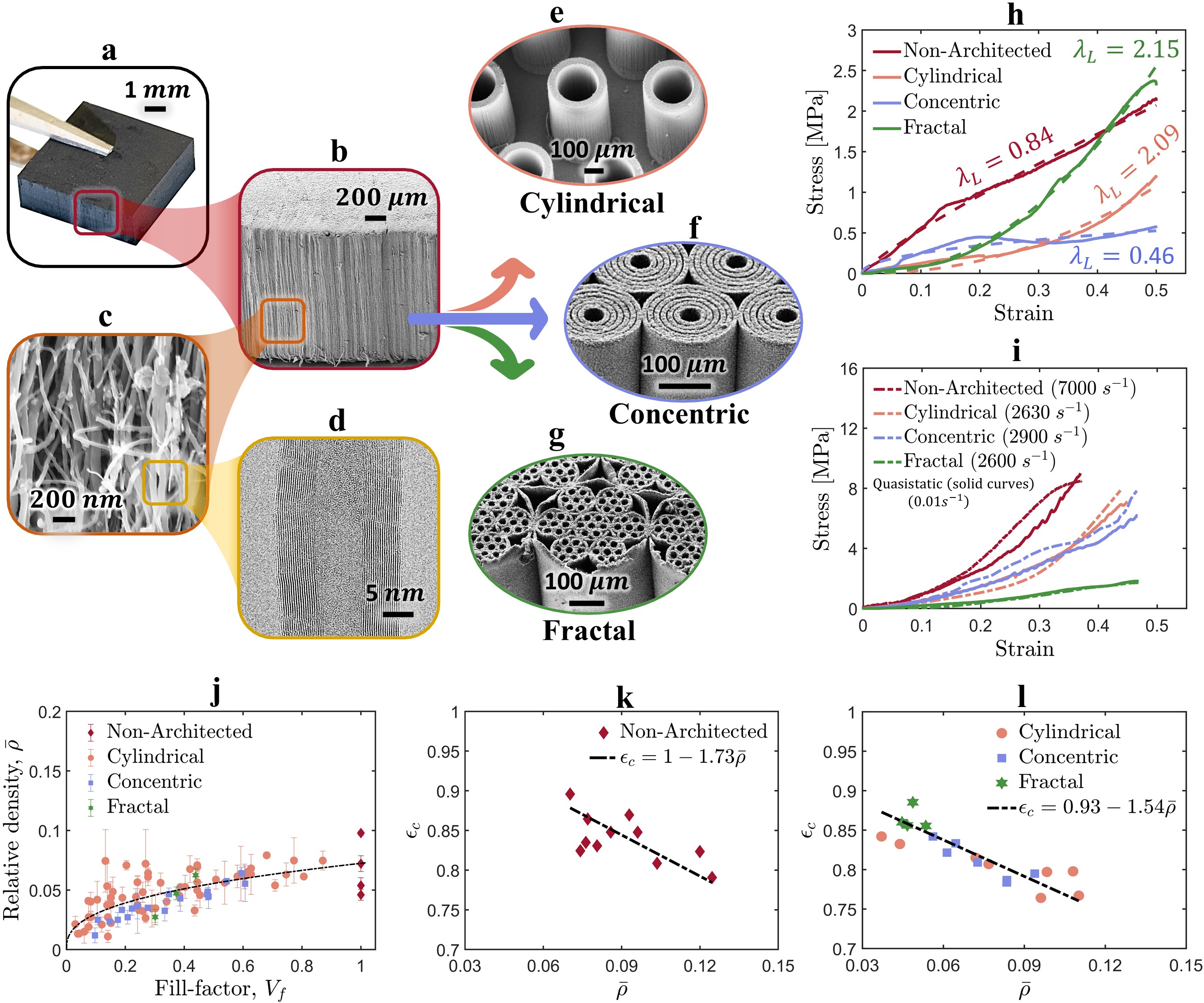}
	\caption{(a,b,c,d) Hierarchical vertically aligned carbon nanotube foam with structural lengthscales spanning from nanometers to millimeters. VACNT foams with mesoscale architecture having hexagonal close-packed arrays of cylinders (e), concentric cylinders (f), and self-similar fractal ($2nd$ order) (g). (h) Representative experimentally measured quasistatic stress-strain responses of non-architected and architected VACNT foams along with the corresponding power law models. (i) Comparison of quasistatic and dynamic stress-strain responses of non-architected and architected VACNT samples measured using commercial load frame and Kolsky bar setup respectively. (j) Relative density of various non-architected and architected VACNT foam samples as a function of the fill-factor of the architecture $(V_f)$. (k) Critical strain of non-architected VACNT foams as a function of relative density. (l) Critical strain of architected VACNT foams as a function of relative density.}
	\label{fig6}
\end{figure}

\section*{Architected VACNT Foams}

The underlying assumption behind the minimization problem we pursued in the previous section was that the material parameters, such as $\bar{\rho}$, $\lambda_L$, and $\alpha$, are independent of each other. In reality, they exhibit some inter-dependency, and not all possible combinations can occur, especially in stochastic polymeric and metallic foams. For example, a nonlinear stress-strain response (where $\lambda_L>1$) is usually observed in foams with large relative densities. In contrast, architected foams offer versatility in the design space and allow independent tunability of different mechanical properties. The vertically aligned carbon nanotube (VACNT) foams with mesoscale architecture are particularly interesting because of their hierarchical structure that can be tailored to achieve broad range of properties. The non-architected VACNT foams themselves exhibit exceptional modulus and energy absorption comparable to metallic foams, while their densities, compressibility, and strain-recovery are similar to polymeric foams \cite{cao2005super}. Moreover, unlike polymeric foams, which exhibit viscoelastic behavior and slowly recover from strain after compression, VACNT foams exhibit fast strain recovery, making them useful for countering repetitive impacts.
These exceptional properties arise from a hierarchical structure with features across various length scales. Multi-walled CNTs (MWCNTs) at the nanoscale (\Cref{fig6}(d)), a random forest of entangled CNTs at the microscale (\Cref{fig6}(c)), and a structure of nominally vertically aligned CNTs at the mesoscale (\Cref{fig6}(b)) all culminate in a monolithic, seemingly solid foam at the macroscale (\Cref{fig6}(a)). 
We introduced an additional level of structural hierarchy at the mesoscale by synthesizing VACNTs on a photo-lithographically prepatterned silicon wafer substrate. We selectively deposit chromium in areas where we do not want CNTs to grow (the inverse of architecture), allowing growth only in the region defined by the architecture \cite{chawla2023disrupting}. \Cref{fig6} illustrates SEM images of three different mesoscale architectures: a hexagonally packed cylindrical architecture (\Cref{fig6}(e)), a concentric cylinder architecture (\Cref{fig6}(f)), and a self-similar fractal architecture (\Cref{fig6}(g)).

In our previous works on cylindrical and concentric-cylinder architected VACNT foams, we reported tunability in density-dependent scaling, specific modulus, and relative density as functions of various architectural parameters \cite{chawla2023disrupting,chawla2022superior}. Furthermore, by increasing the gap between cylinders and thereby reducing lateral interactions between them, we demonstrated a transformation in the shape of the constitutive stress-strain curve from nonlinear to sublinear (see \Cref{fig6}(h)).
Here, we introduce a self-similar fractal architecture that enhances interconnectivity between mesoscale cylinders at much lower relative densities, enabling higher specific energy absorption. We investigate fractal-architected samples and compare their properties with our previous cylindrical and concentric VACNT samples to identify optimal architectures for compact and lightweight shock absorbers.

\begin{figure}[ht!]
	\centering
	\includegraphics[width=\textwidth]{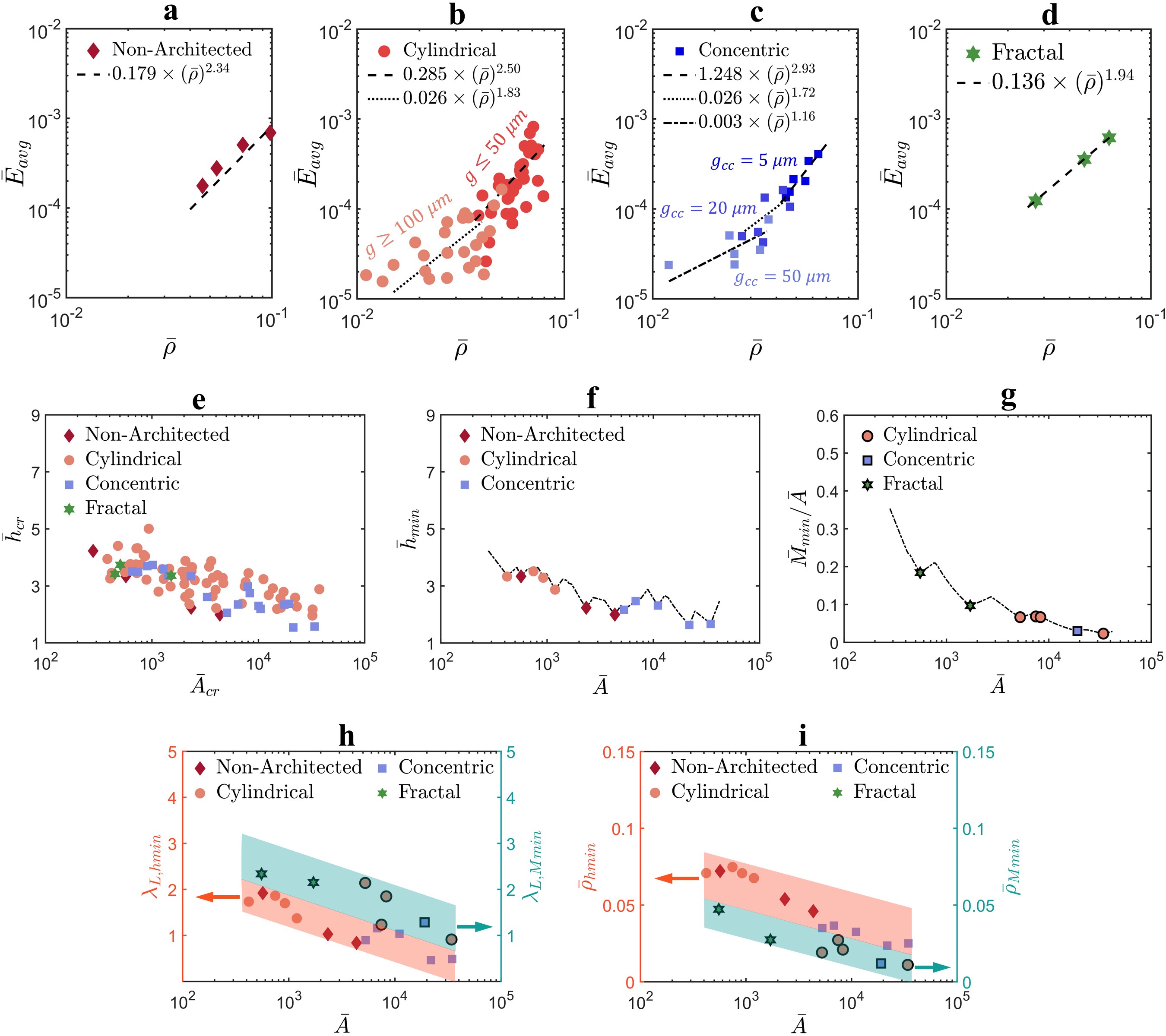}
	\caption{Scaling of normalized average modulus with relative density of non-architected (a), cylindrical (b), concentric (c), and fractal architectures (d). (e) Critical thickness vs. critical area of all VACNT foams. Minimum thickness (f) and minimum mass per unit area (g) as a function of dimensionless area. (h) $\lambda_{L,hmin}$ compared to $\lambda_{L,Mmin}$. (i) $\bar{\rho}_{hmin}$ compared to $\bar{\rho}_{Mmin}$}
	\label{fig7}
\end{figure}

\Cref{fig6}(h) shows representative experimentally measured quasistatic (strain rate of $0.01\;s^{-1}$) stress-strain responses of non-architected VACNT foam as well as architected foams, including the corresponding power law models (dashed curves). As shown, architected VACNT foams enhances the range of values of $\lambda_L$, which was limited to nearly linear $\lambda_L\approx1$ for non-architected foams, to sublinear $(\lambda_L<1)$ and nonlinear $(\lambda_L>1)$. The shape of the stress-strain curve or the value of $\lambda_L$ depends on the specific architectural parameters that elicit specific deformation mechanism of cylinders within the architecture. For example, in concentric cylinder architecture (\Cref{fig6}(f)), a column buckling of cylinders with larger gap between the concentric cylinders results in a sublinear stress-strain curve, whereas a progressive shell buckling of cylinders with smaller gap results in a nonlinear stiffening stress-strain response \cite{chawla2023disrupting}. Moreover, the stress-strain responses of both architected and non-architected VACNT foams are strain rate independent (\Cref{fig6}(i)) up to very large strain rates \cite{raney2013rate,lattanzi2015dynamic,xu2010carbon}. In \Cref{fig6}, the stress-strain responses of a cylindrically architected VACNT foam we tested at quasistatic and dynamic strain rates, reveals almost no effect of strain rate. This strain rate independence and the richness of material parameter space makes VACNT foams an ideal material system for us to demonstrate the effectiveness of our model in real materials.

We synthesized multiple samples in each architecture category by varying the dimensions of architectural features. For example, by varying the inner diameter $(D_{in})$, wall-thickness $(t_w)$, and gap $(g)$ between the cylinders in cylindrically architected foams, we synthesized 60 different types of samples \cite{chawla2022superior}. In concentric architecture, we varied the inner gap and the number of rings of concentric cylinders, resulting in a total of 18 samples \cite{chawla2023disrupting}. For fractal architecture, we synthesized 3 types of samples with different orders of self-similarity (see Figure S5, S6 in \hyperref[section:sd]{SI}). Moreover, we repeated the synthesis of each kind of sample three times and averaged the mechanical properties to account for any variability in the samples of a given architecture.

\begin{figure}[ht!]
	\centering
	\includegraphics[width=\textwidth]{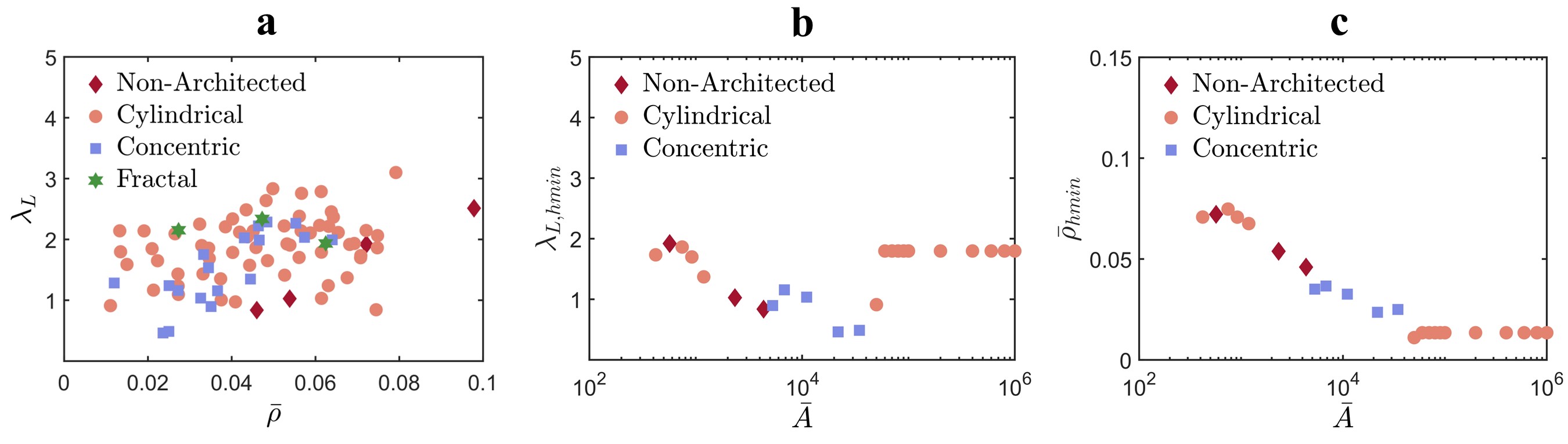}
	\caption{(a) $\lambda_L$ vs. $\bar{\rho}$ for VACNT foams. (b) $\lambda_L$ corresponding to minimum thickness as a function of area for VACNT foams. (c) $\bar{\rho}$ corresponding to minimum thickness as a function of area for VACNT foams.}
	\label{fig9b}
\end{figure}

We calculated the relative density $(\bar{\rho})$ by dividing the bulk density $(\rho)$ of VACNT foams by the density of highly packed CNTs $(\rho_s\sim2.26 \;g/cm^3)$ \cite{jung2018can}. In \Cref{fig6}(j), we plotted the relative densities of all VACNT foams as a function of the fill-factor $(V_f)$ of the architectures (see \hyperref[section:sd]{SI} for derived expressions of $V_f$ for different architectures). It is evident that the relative density is adjustable as a function of fill-factor and exhibits a wide range across samples, from $\bar{\rho}=0.01$ to $\bar{\rho}=0.1$. As the fill-factor approaches $1$, the relative density of architected VACNT foams asymptotically approaches the relative density of non-architected VACNT foams. Similar to polymeric foams in \Cref{fig2}(b), we measured the critical strain ${\epsilon_c}$ of both non-architected and architected VACNT foams using the energy absorption efficiency method \cite{li2006compressive}. For both types of foams, the critical strain $\epsilon_c$ varies linearly as a function of $\bar{\rho}$ with a slope of $\sim 2$, which is typically observed in crushable metallic foams \cite{gibson2003cellular}. 

We observe a wide range of density dependent scaling of elastic modulus among different architectures. In \Cref{fig7}(a,b,c,d), we plot average relative modulus $(\bar{E}_{avg})$ of VACNT foams as a function of relative density, where $\bar{E}_{avg}$ is a dimensionless modulus calculated from the experimentally measured stress-strain response of VACNT foams. $\bar{E}_{avg}$ is a measure of the elastic energy stored in the foam during compression.  Unlike $E_L$, $\bar{E}_{avg}$ is normalized by $\lambda_L+1$, thus it scales more consistently with relative density (see \hyperref[section:sd]{SI} for details). As a function of loading modulus $E_L$ and $\lambda_L$ obtained from power-law fits, the expression of $\bar{E}_{avg}$ is given as follows

\begin{equation}
    \bar{E}_{avg} = {E_L \over E_s} \times {(\lambda_L+1)^{\beta-1}} {\epsilon_{p}^{\lambda_L-1}} \times (2-\delta)
\end{equation}

where, $E_s$ is the elastic modulus of highly densely packed CNTs ($E_s\sim 15\;GPa$) \cite{pathak2009viscoelasticity}, $\epsilon_{p}$ is the peak compression strain applied while measuring the stress-strain response, and $\delta$ is the damping capacity---a ratio of hysteretic energy dissipated in the loading-unloading cycle divided by the area under the loading curve \cite{chawla2023disrupting}. For non-architected and fractal-architected, the scaling is nonlinear with scaling exponent $\alpha=2.34$ and $\alpha=1.94$ respectively. However, for cylindrically-architected and concentric cylinder architectures, the scaling is tunable as a function of the external gap between the cylinders and the internal gap between the concentric cylinders respectively (see Figure S5, S6 in \hyperref[section:sd]{SI}). 

Using the measured material parameters $(\bar{\rho}, \lambda_L, E_L)$ we calculated dimensionless critical thickness and critical area which are plotted in \Cref{fig7}(e). To identify the VACNT foam samples that will result in minimum thickness, we implemented the framework that we demonstrated earlier in \Cref{fig4}(b). In the thickness minimization process, starting with $85$ samples, we were able to condense down to $12$ samples that will result in minimum thickness for a given area $(\bar{A})$ lying in the range shown in \Cref{fig7}(f). Similarly, we identified VACNT foam samples that will result in minimum mass for a given $\bar{A}$. For minimum mass, we were able to condense down to $7$ samples which all turned out to be architected VACNT foams, because of the comparatively large relative densities associated with non-architected VACNT foams. In \Cref{fig7}(h), we compare $\lambda_L$ corresponding to minimum thickness $(\lambda_{L,hmin})$ with minimum mass $(\lambda_{L,Mmin})$. The data points corresponding to $\lambda_{L,Mmin}$ are plotted with black outline to distinguish from $\lambda_{L,hmin}$. The data points seem to form distinctive bands with $\lambda_{L,hmin}$ forming a band of lower overall values compared to $\lambda_{L,Mmin}$. In contrast, for relative density, $\bar{\rho}_{Mmin}$ values form a band of lower overall values compared to $\bar{\rho}_{hmin}$ (\Cref{fig7}(i)). In summary, a higher $\bar{\rho}$ and a sublinear $\lambda_L$ are favourable to achieve minimum thickness whereas a smaller $\bar{\rho}$ and a nonlinear $\lambda_L$ are needed to achieve minimum mass. 

As a function of $\bar{A}$, both $\lambda_{L,hmin}$ and $\bar{\rho}_{hmin}$ exhibit a decreasing trend, which differs from the behavior observed in \Cref{fig4}. Previously, we assumed that material parameters were independent, meaning that for each value of $\bar{\rho}$ within the range $0.01 < \bar{\rho} < 0.1$, any value of $\lambda_L$ within $0.1 < \lambda_L < 3$ could be paired with it. In VACNT foams, however, there is a weak dependence between $\lambda_L$ and $\bar{\rho}$, where $\lambda_L$ scales weakly with $\bar{\rho}$ (\Cref{fig9b}(a)). For samples with higher $\bar{\rho}$, $\bar{A}_{cr}$ is smaller because $\bar{A}_{cr} \propto 1/\bar{\rho}^{\alpha}$, but $\bar{h}_{cr}$ is larger since $\bar{h}_{cr} \propto \lambda_L+1$, leading to a decreasing trend in $\bar{h}_{cr}$ with increasing $\bar{A}_{cr}$ (\Cref{fig7}(e)). This trend is also reflected in the plot of $\lambda_{L,hmin}$. Notably, the decreasing trend persists only within the $\bar{A}$ range shown in \Cref{fig7}(h), which corresponds to the $\bar{A}_{cr}$ values in \Cref{fig7}(e). For higher $\bar{A}$ values $(\bar{A}>10^5)$, $\lambda_{L,hmin}$ increases and then stabilizes (\Cref{fig9b}(b)), while $\bar{\rho}_{hmin}$ decreases before becoming constant (\Cref{fig9b}(c)), similar to \Cref{fig4}(d,e). Therefore, our finding that foams with nonlinear stress-strain responses perform better at higher $\bar{A}$ values holds true for VACNT foams as well (see Figure S11 in \hyperref[section:sd]{SI}). $\lambda_L$ and $\bar{\rho}$ corresponding to minimum mass i.e., $\lambda_{L,Mmin}$ and $\bar{\rho}_{Mmin}$ also follow similar trends for extended ranges of $\bar{A}$ (see Figure S10 in \hyperref[section:sd]{SI}).

\begin{figure}[t]
	\centering
	\includegraphics[width=\textwidth]{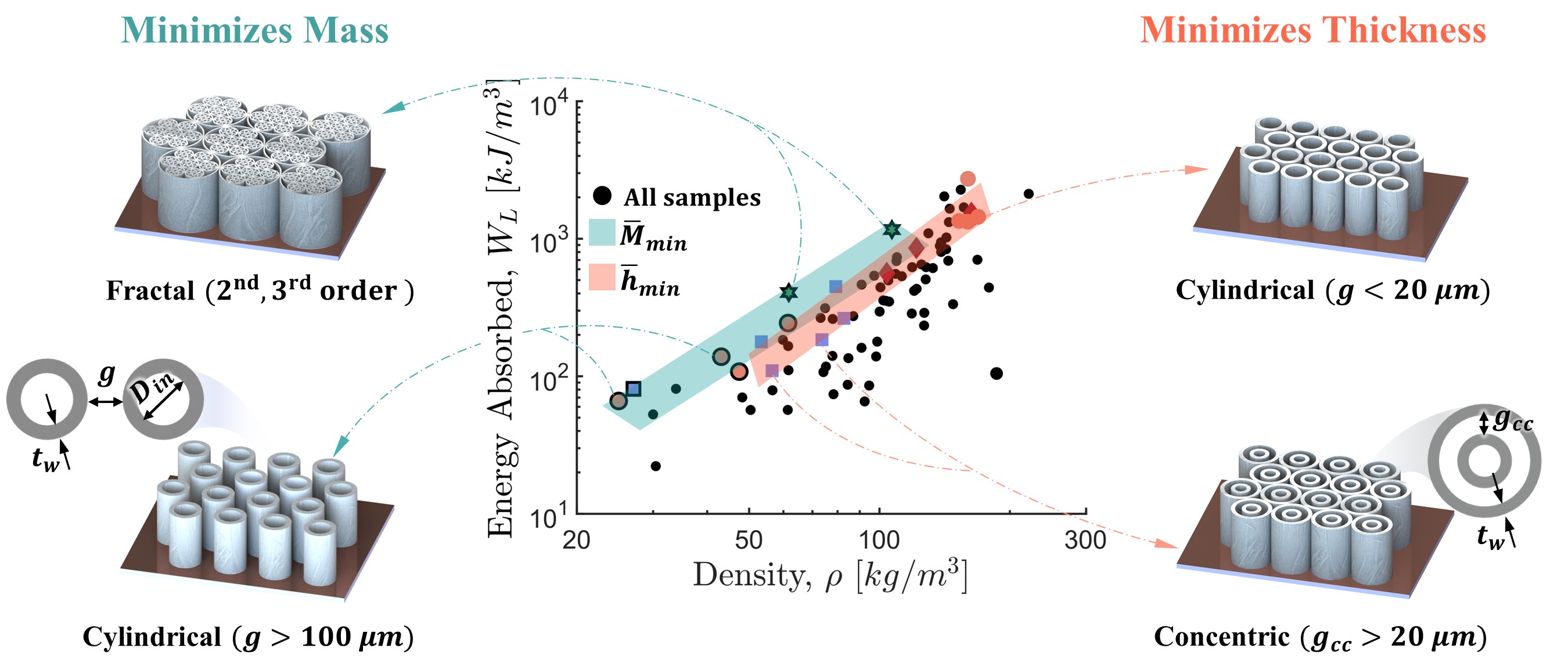}
	\caption{Volumetric energy absorption as a function of density for all the VACNT foam samples. The data points for samples that minimizes mass are shown with black outline. The illustration of the architectures of different VACNT samples that outperform others are also shown}
	\label{fig8}
\end{figure}

In \Cref{fig8}, we present a plot showing volumetric energy absorption $(W_L)$ versus density, where we have isolated the best-performing VACNT samples from a total of 85 samples. The samples that minimize thickness ($\bar{h}_{min}$) are represented by colored markers, while data points for samples that minimize mass ($\bar{M}_{min}$) are depicted with the same markers but with black borders. All other samples are shown using solid black dots. Energy absorption scales with density, and since $\lambda_L$ exhibits a weak correlation with $\bar{\rho}$ (\Cref{fig9b}(a)), nonlinear foams demonstrate slightly higher energy absorption for a given $\bar{\rho}$. However, these results involve VACNT samples with varying scaling factors $\alpha$. The precise relationship between energy absorption and $\lambda_L$ for constant $\alpha$ and $\bar{\rho}$ cannot be determined due to finite number of VACNT samples and interdependence of $\alpha$, $\lambda_L$, and $\bar{\rho}$ (\Cref{fig7}(a,b,c,d)).  

The samples corresponding to $\bar{M}_{min}$ and $\bar{h}_{min}$ form two distinct bands in the top region of the scatter plot, demonstrating maximum energy absorption for a given density. This phenomenon can be clearly understood by examining the equation for energy absorption per unit volume. Using \cref{eq15} and \cref{eq16}, we can restructure the equation for energy absorption \cref{eq3_2} as follows

\begin{equation}
    \frac{W_L}{E_s}=\frac{1}{\bar{h}_{c r}} \times \frac{1}{\bar{A}_{c r}} \;\;\rightarrow \;\; \bar{h}_{c r} \bar{A}_{c r}=\frac{E_s}{W_L}
    \label{eq25}
\end{equation}

where $\bar{h}_{c r} \bar{A}_{c r}$ is the critical volume. The expression above indicates that foams with higher $W_L$ will result in lower critical volume therefor lower thickness for a given area, as observed in \Cref{fig8}. On multiplying both sides of the \cref{eq25} with $\bar{\rho}$, we get

\begin{equation}
    \bar{\rho} \bar{h}_{c r} \bar{A}_{c r}=\frac{E_s}{W_L} \times \bar{\rho} \;\;\rightarrow \;\;\bar{M}_{c r}=\frac{E_s}{W_L} \times \bar{\rho}
\end{equation}

where $\bar{M}_{cr}$ represents the critical mass, which minimizes for higher $W_L$ and lower relative density, as observed in \Cref{fig8}. This analysis demonstrates that in addition to the material's property parameters $(\lambda_L,\bar{\rho},\alpha)$, exhibiting a high energy absorption per unit volume is also crucial in designing thin and lightweight shock absorbers.

The architectures corresponding to the majority of samples lying in these bands are illustrated along with their geometric design parameters. For $\bar{M}_{min}$, the samples include higher-order fractal ($2nd$ and $3rd$ order) and sparsely packed cylindrically architected samples (with $g>100\; \mu m$) due to their low densities and high energy absorption capabilities. On the other hand, samples associated with $\bar{h}_{min}$ include tightly packed cylindrical architected and non-architected samples due to their highest energy absorption. Additionally, sparsely $(g_{cc}>20\mu m)$ packed concentric cylindrically architected foams are present due to their sublinear stress-strain response (\Cref{fig7}(h)) and consequently lower critical thickness ($\bar{h}_{cr}$) (see \Cref{eq15}). While a lot of samples, other than the optimal ones, lying in the $\bar{h}_{min}$ band exhibit considerably high energy absorption at a given density, their $\lambda_L$ values are too large to minimize thickness. 

In our theoretical calculations, we achieved simultaneously minimized thickness and mass for a large range of $\lambda_L$ and $\bar{\rho}$ varying independently. We showed that for $\alpha\geq 1.5$, thickness and mass minimizes simultaneously. For VACNT foams, we witnessed similar behavior when both thickness and mass were minimized across a set of samples following a particular fixed scaling, $\alpha$, (see Figure S12 in \hyperref[section:sd]{SI}) like we did in theoretical analysis. However, when considering all VACNT foams collectively, gaps in the material parameter space lead to variations: some foams minimize thickness, while others minimize mass. For example, the linear density-dependent scaling observed in concentric VACNT foams for $g_{cc}=50\;\mu m$ (\Cref{fig7}(c)) occurs over a narrow range of low relative densities ($0.012<\bar{\rho}<0.037$), whereas cubic scaling for $g_{cc}=5\;\mu m$ is seen at higher relative densities ($0.044<\bar{\rho}<0.064$). As a result, lower $\alpha$ and higher $\bar{\rho}$ become conflicting, deviating from the parameter independence assumed in our theoretical results. A similar pattern is evident in cylindrically architected VACNT foams (\Cref{fig7}(b)). Moreover there is a slight interdependence between $\lambda_L$ and $\bar{\rho}$ (\Cref{fig9b}(a)) as well, whereas we also assumed them to be independent in theoretical analysis. Nevertheless, this study demonstrates that, despite some gaps in the parameter space, VACNT foams offer a rich material property landscape with a wide range of parameters suitable for designing efficient shock absorbers. We envision that using novel hierarchical graph network based machine-learning methods to design architectures can potentially result in a broadened parameter space, allowing for independent tunability of $\lambda_L$, $\alpha$, and $\bar{\rho}$. Such independent tunability can possibly lead to simultaneous thickness and mass minimization in various types of architectures.

\section*{Conclusion}

In conclusion, our study highlights the pivotal role of a foam pad's geometry in addition to the constitutive behavior of the material in determining the mechanical performance of the protective layer. By employing a simplified kinematic model and dimensional analysis, we established constraints on the cross-sectional area and thickness of foam pads with known mechanical properties. These constraints ensure that both the maximum compressive strain within the foam and the maximum acceleration experienced by the protected object remain below desired limits. Using these constraints as a design framework, we identified the mechanical properties necessary for achieving the thinnest, lightest, or a combination of both in foam pads for a given cross-sectional area.
Contrary to prevailing beliefs that a plateau in a stress-strain response is essential for all shock absorbing applications, our findings suggest that foams with nonlinear responses, with gradually increasing stress in strain, can enable the design of both thin and lightweight foam pads in various extreme protective applications. More specifically, we found that foams with a nonlinear stress-strain response can outperform foams exhibiting a plateau-type stress-strain response when protective foam pads of large area coverage is required, such as in helmet liners and packaging applications.
Additionally, we also discover that foam materials with nonlinear density-dependent scaling of modulus can lead to simultaneously lightweight and thin protective energy absorbers.
Using our non-dimensional design framework, we demonstrate optimal designs in hierarchically architected VACNT foams with cylindrical, concentric cylindrical, and fractal architectures.
Our generalized design framework can be applied to any architected and stochastic foams with known mechanical properties to design compact and lightweight energy absorbing pads for diverse protective applications by identifying the best-performing architectures from a multitude of options.

\section*{Methods}

\subsection*{Synthesis of architected VACNT foams}

To synthesize architected VACNT foams, we utilize silicon wafer substrates with predefined microscale patterns created using photolithography. First, we spin-coat a $10 \;\mu m$ thick layer of photoresist (MICROPOSIT S1813) on a $100\;mm$ diameter p-type silicon wafer ($100$ crystal orientation and $500\;\mu m$ thickness) at $3000\;rpm$ for $30\;s$ and prebake it on a hot plate at $383\;K$ for $45\;s$ to remove any solvents. After spin-coating, we partially dice the wafer up to $30\;\%$ of its thickness into $5\times5\;mm$ squares and expose it to ultraviolet light through a chrome/soda-lime photomask with predefined micropatterns (manufactured by Photo Sciences, Valencia, CA, USA). After exposure to $405\;nm$ UV light at $80\;mJ/cm^2$ for 8 seconds, we remove the unexposed photoresist in an MF321 developer bath for $30\;s$ and then coat the wafer with a $20\;nm$ thin film of chromium at $0.05\;nm/s$ using a metal evaporator. To remove the remaining photoresist (UV exposed), we place the wafer in an acetone bath for $30\;s$, thereby leaving only chromium on the surface with the predefined architecture.

\subsection*{Mechanical characterization}

To measure the quasistatic stress-strain responses of VACNT foams and polymeric foams reported in this article, we performed ramp compression experiments using a commercial load frame, the Instron Electropulse E3000, equipped with a 5kN load cell. We compressed the samples at constant strain rate in a ramp waveform up to a strain exceeding the onset of densification strain to measure the critical strain and then unloaded at the same strain rate. To measure energy absorption and critical strain as functions of strain rate in polymeric foams, we varied the strain rates from $0.001\;s^{-1}$ to $0.1\;s^{-1}$. To characterize the mass density, we measured the sample's mass with a microbalance, measured its thickness with a micrometer gauge, and calculated its volume using the sample's area $(5\times 5\;mm^2)$ and thickness.

To measure the high strain rate response of both VACNT and polymeric foam materials, we built a tabletop Kolsky bar apparatus equipped with 3/8 inch (9.52 mm) diameter high-strength 2024 aluminum alloy bars. The incident bar and transmitter bar measured 700 mm and 550 mm in length, respectively. We used striker bars made of the same material with a 1/2 inch (12.7 mm) diameter and lengths varying from 80 mm to 120 mm. We mounted resistive strain gauges from Micro-Measurements (CEA-13-250UN-350) on the incident bar, 350 mm away from the sample-bar interface. To detect the weak transmitted signal, we mounted semiconductor strain gauges with high signal to noise ratio from KYOWA (KSPB-1-350-E4) on the transmitter bar, 265 mm away from the sample-bar interface. An air compressor-powered gas gun launched the striker at adjustable velocities, achieving strain rates between $2000\;s^{-1}$ to $7000\;s^{-1}$.

\section*{Acknowledgement}

This research is supported by the U.S. Office of Naval Research under PANTHER program (award number N000142112044) through Dr. Timothy Bentley as well as the by the solid mechanics program of the U.S. Army Research Office (award number: W911NF2010160) through Dr. Denise Ford. The authors acknowledge the use of facilities and instrumentation at the Wisconsin Centers for Nanoscale Technology (WCNT) partially supported by the NSF through the University of Wisconsin Materials Research Science and Engineering Center (DMR-1720415). We also acknowledge Team Wendy for their Zorbium soft and hard polyurethane foams that were compared with our VACNT foams for the mechanical performance.

\section*{Author Contributions}

A.G. performed the theoretical analysis, performed the experiments on polymeric foams, analyzed the data, and prepared the figures. K.C. synthesized the VACNT foam samples and performed the quasistatic compression testing. B.M. designed, built, and performed the Kolsky bar experiments. D.S. conducted the finite-element simulations on architected lattice. R.T. conceived and supervised the project. A.G. and R.T. wrote the paper with input from all the authors.

\section*{Competing Interests}

The authors declare no competing interest.

\section*{Supplementary Information}

Supplementary material related to this article can be found online
\label{section:sd}

\section*{Code availability}

Our MATLAB script for designing thin and lightweight shock absorber from a given set of available foam materials and known system parameters is available on \href{https://github.com/ThevamaranLab/Shock-Absorber-Design.git}{https://github.com/ThevamaranLab/Shock-Absorber-Design.git}

\bibliographystyle{unsrt}
\bibliography{citations.bib}

\begin{thebibliography}{10}

\bibitem{ramirez2018evaluation}
BJ~Ramirez and V~Gupta.
\newblock Evaluation of novel temperature-stable viscoelastic polyurea foams as helmet liner materials.
\newblock {\em Materials \& Design}, 137:298--304, 2018.

\bibitem{shuaeib2007new}
FM~Shuaeib, AMS Hamouda, SV~Wong, RS~Radin Umar, and MMH~Megat Ahmed.
\newblock A new motorcycle helmet liner material: The finite element simulation and design of experiment optimization.
\newblock {\em Materials \& design}, 28(1):182--195, 2007.

\bibitem{zhang1994mechanical}
J~Zhang and MF~Ashby.
\newblock Mechanical selection of foams and honeycombs used for packaging and energy absorption.
\newblock {\em Journal of Materials Science}, 29:157--163, 1994.

\bibitem{li2007dynamic}
Ke~Li, X-L Gao, and Jun Wang.
\newblock Dynamic crushing behavior of honeycomb structures with irregular cell shapes and non-uniform cell wall thickness.
\newblock {\em International Journal of Solids and Structures}, 44(14-15):5003--5026, 2007.

\bibitem{barnes2014dynamic}
AT~Barnes, K~Ravi-Chandar, S~Kyriakides, and S~Gaitanaros.
\newblock Dynamic crushing of aluminum foams: Part i--experiments.
\newblock {\em International Journal of Solids and Structures}, 51(9):1631--1645, 2014.

\bibitem{li2011crashworthiness}
Meng Li, Zongquan Deng, Rongqiang Liu, and Hongwei Guo.
\newblock Crashworthiness design optimisation of metal honeycomb energy absorber used in lunar lander.
\newblock {\em International Journal of crashworthiness}, 16(4):411--419, 2011.

\bibitem{gibson2003cellular}
Lorna~J Gibson.
\newblock Cellular solids.
\newblock {\em MRS Bulletin}, 28(4):270--274, 2003.

\bibitem{schaedler2016architected}
Tobias~A Schaedler and William~B Carter.
\newblock Architected cellular materials.
\newblock {\em Annual Review of Materials Research}, 46:187--210, 2016.

\bibitem{bauer2016approaching}
Jens Bauer, Almut Schroer, Ruth Schwaiger, and Oliver Kraft.
\newblock Approaching theoretical strength in glassy carbon nanolattices.
\newblock {\em Nature Materials}, 15(4):438--443, 2016.

\bibitem{chen2019stiff}
Wen Chen, Seth Watts, Julie~A Jackson, William~L Smith, Daniel~A Tortorelli, and Christopher~M Spadaccini.
\newblock Stiff isotropic lattices beyond the maxwell criterion.
\newblock {\em Science Advances}, 5(9):eaaw1937, 2019.

\bibitem{crook2020plate}
Cameron Crook, Jens Bauer, Anna Guell~Izard, Cristine Santos~de Oliveira, Juliana Martins de Souza~e Silva, Jonathan~B Berger, and Lorenzo Valdevit.
\newblock Plate-nanolattices at the theoretical limit of stiffness and strength.
\newblock {\em Nature Communications}, 11(1):1579, 2020.

\bibitem{zheng2023unifying}
Li~Zheng, Konstantinos Karapiperis, Siddhant Kumar, and Dennis~M Kochmann.
\newblock Unifying the design space and optimizing linear and nonlinear truss metamaterials by generative modeling.
\newblock {\em Nature Communications}, 14(1):7563, 2023.

\bibitem{berger2017mechanical}
JB~Berger, HNG Wadley, and RM~McMeeking.
\newblock Mechanical metamaterials at the theoretical limit of isotropic elastic stiffness.
\newblock {\em Nature}, 543(7646):533--537, 2017.

\bibitem{schaedler2011ultralight}
Tobias~A Schaedler, Alan~J Jacobsen, Anna Torrents, Adam~E Sorensen, Jie Lian, Julia~R Greer, Lorenzo Valdevit, and Wiliam~B Carter.
\newblock Ultralight metallic microlattices.
\newblock {\em Science}, 334(6058):962--965, 2011.

\bibitem{zheng2014ultralight}
Xiaoyu Zheng, Howon Lee, Todd~H Weisgraber, Maxim Shusteff, Joshua DeOtte, Eric~B Duoss, Joshua~D Kuntz, Monika~M Biener, Qi~Ge, Julie~A Jackson, et~al.
\newblock Ultralight, ultrastiff mechanical metamaterials.
\newblock {\em Science}, 344(6190):1373--1377, 2014.

\bibitem{wang2022achieving}
Yujia Wang, Xuan Zhang, Zihe Li, Huajian Gao, and Xiaoyan Li.
\newblock Achieving the theoretical limit of strength in shell-based carbon nanolattices.
\newblock {\em Proceedings of the National Academy of Sciences}, 119(34):e2119536119, 2022.

\bibitem{portela2021supersonic}
Carlos~M Portela, Bryce~W Edwards, David Veysset, Yuchen Sun, Keith~A Nelson, Dennis~M Kochmann, and Julia~R Greer.
\newblock Supersonic impact resilience of nanoarchitected carbon.
\newblock {\em Nature Materials}, 20(11):1491--1497, 2021.

\bibitem{butruille2024decoupling}
Thomas Butruille, Joshua~C Crone, and Carlos~M Portela.
\newblock Decoupling particle-impact dissipation mechanisms in 3d architected materials.
\newblock {\em Proceedings of the National Academy of Sciences}, 121(6):e2313962121, 2024.

\bibitem{suhr2007fatigue}
Jonghwan Suhr, Pushparaj Victor, Lijie Ci, Subbalakshmi Sreekala, Xianfeng Zhang, Omkaram Nalamasu, and Pulickel~M Ajayan.
\newblock Fatigue resistance of aligned carbon nanotube arrays under cyclic compression.
\newblock {\em Nature Nanotechnology}, 2(7):417--421, 2007.

\bibitem{meza2015resilient}
Lucas~R Meza, Alex~J Zelhofer, Nigel Clarke, Arturo~J Mateos, Dennis~M Kochmann, and Julia~R Greer.
\newblock Resilient 3d hierarchical architected metamaterials.
\newblock {\em Proceedings of the National Academy of Sciences}, 112(37):11502--11507, 2015.

\bibitem{moestopo2023knots}
Widianto~P Moestopo, Sammy Shaker, Weiting Deng, and Julia~R Greer.
\newblock Knots are not for naught: Design, properties, and topology of hierarchical intertwined microarchitected materials.
\newblock {\em Science Advances}, 9(10):eade6725, 2023.

\bibitem{abayazid2024viscoelastic}
Fady~F Abayazid and Mazdak Ghajari.
\newblock Viscoelastic circular cell honeycomb helmet liners for reducing head rotation and brain strain in oblique impacts.
\newblock {\em Materials \& Design}, page 112748, 2024.

\bibitem{chawla2023disrupting}
Komal Chawla, Abhishek Gupta, and Ramathasan Thevamaran.
\newblock Disrupting density-dependent property scaling in hierarchically architected foams.
\newblock {\em ACS nano}, 17(11):10452--10461, 2023.

\bibitem{ashby2006properties}
Michael~F Ashby.
\newblock The properties of foams and lattices.
\newblock {\em Philosophical Transactions of the Royal Society A: Mathematical, Physical and Engineering Sciences}, 364(1838):15--30, 2006.

\bibitem{cao2005super}
Anyuan Cao, Pamela~L Dickrell, W~Gregory Sawyer, Mehrdad~N Ghasemi-Nejhad, and Pulickel~M Ajayan.
\newblock Super-compressible foamlike carbon nanotube films.
\newblock {\em Science}, 310(5752):1307--1310, 2005.

\bibitem{thevamaran2015shock}
Ramathasan Thevamaran, Eric~R Meshot, and Chiara Daraio.
\newblock Shock formation and rate effects in impacted carbon nanotube foams.
\newblock {\em Carbon}, 84:390--398, 2015.

\bibitem{chawla2024superior}
Komal Chawla, Jizhe Cai, Dakotah Thompson, and Ramathasan Thevamaran.
\newblock Superior thermal transport properties of vertically aligned carbon nanotubes tailored through mesoscale architectures.
\newblock {\em Carbon}, 216:118526, 2024.

\bibitem{xu2010carbon}
Ming Xu, Don~N Futaba, Takeo Yamada, Motoo Yumura, and Kenji Hata.
\newblock Carbon nanotubes with temperature-invariant viscoelasticity from--196 to 1000 c.
\newblock {\em Science}, 330(6009):1364--1368, 2010.

\bibitem{raney2013rate}
Jordan~R Raney, Fernando Fraternali, and C~Daraio.
\newblock Rate-independent dissipation and loading direction effects in compressed carbon nanotube arrays.
\newblock {\em Nanotechnology}, 24(25):255707, 2013.

\bibitem{lattanzi2014geometry}
Ludovica Lattanzi, Luigi De~Nardo, Jordan~R Raney, and Chiara Daraio.
\newblock Geometry-induced mechanical properties of carbon nanotube foams.
\newblock {\em Advanced Engineering Materials}, 16(8):1026--1031, 2014.

\bibitem{chawla2022superior}
Komal Chawla, Abhishek Gupta, Abhijeet~S Bhardwaj, and Ramathasan Thevamaran.
\newblock Superior mechanical properties by exploiting size-effects and multiscale interactions in hierarchically architected foams.
\newblock {\em Extreme Mechanics Letters}, 57:101899, 2022.

\bibitem{ha2023rapid}
Chan~Soo Ha, Desheng Yao, Zhenpeng Xu, Chenang Liu, Han Liu, Daniel Elkins, Matthew Kile, Vikram Deshpande, Zhenyu Kong, Mathieu Bauchy, et~al.
\newblock Rapid inverse design of metamaterials based on prescribed mechanical behavior through machine learning.
\newblock {\em Nature Communications}, 14(1):5765, 2023.

\bibitem{liu2022growth}
Ke~Liu, Rachel Sun, and Chiara Daraio.
\newblock Growth rules for irregular architected materials with programmable properties.
\newblock {\em Science}, 377(6609):975--981, 2022.

\bibitem{rusch1970load}
KC~Rusch.
\newblock Load-compression behavior of brittle foams.
\newblock {\em Journal of Applied Polymer Science}, 14(5):1263--1276, 1970.

\bibitem{hassani2012production}
Amir Hassani, Ali Habibolahzadeh, and Hassan Bafti.
\newblock Production of graded aluminum foams via powder space holder technique.
\newblock {\em Materials \& Design}, 40:510--515, 2012.

\bibitem{li2006compressive}
QM~Li, I~Magkiriadis, and John~J Harrigan.
\newblock Compressive strain at the onset of densification of cellular solids.
\newblock {\em Journal of cellular plastics}, 42(5):371--392, 2006.

\bibitem{greer2019three}
Julia~R Greer and Vikram~S Deshpande.
\newblock Three-dimensional architected materials and structures: Design, fabrication, and mechanical behavior.
\newblock {\em MRS Bulletin}, 44(10):750--757, 2019.

\bibitem{cousins1976theory}
RR~Cousins.
\newblock A theory for the impact behavior of rate-dependent padding materials.
\newblock {\em Journal of applied polymer science}, 20(10):2893--2903, 1976.

\bibitem{bhagavathula2022density}
Kapil~Bharadwaj Bhagavathula, Christopher~S Meredith, Simon Ouellet, Sikhanda~S Satapathy, Dan~L Romanyk, and James~D Hogan.
\newblock Density, microstructure, and strain-rate effects on the compressive response of polyurethane foams.
\newblock {\em Experimental Mechanics}, pages 1--15, 2022.

\bibitem{ouellet2006compressive}
Simon Ouellet, Duane Cronin, and Michael Worswick.
\newblock Compressive response of polymeric foams under quasi-static, medium and high strain rate conditions.
\newblock {\em Polymer testing}, 25(6):731--743, 2006.

\bibitem{carlsen2021quantitative}
Rika~Wright Carlsen, Alice~Lux Fawzi, Yang Wan, Haneesh Kesari, and Christian Franck.
\newblock A quantitative relationship between rotational head kinematics and brain tissue strain from a 2-d parametric finite element analysis.
\newblock {\em Brain Multiphysics}, 2:100024, 2021.

\bibitem{mustin1968theory}
Gordon~S Mustin.
\newblock {\em Theory and practice of cushion design}.
\newblock Shock and Vibration Information Center, US Department of Defense, 1968.

\bibitem{worsley2009mechanically}
Marcus~A Worsley, Sergei~O Kucheyev, Joe~H Satcher, Alex~V Hamza, and Theodore~F Baumann.
\newblock Mechanically robust and electrically conductive carbon nanotube foams.
\newblock {\em Applied Physics Letters}, 94(7), 2009.

\bibitem{gross1997elastic}
Joachim Gross, George~W Scherer, Cynthia~T Alviso, and Richard~W Pekala.
\newblock Elastic properties of crosslinked resorcinol-formaldehyde gels and aerogels.
\newblock {\em Journal of non-crystalline solids}, 211(1-2):132--142, 1997.

\bibitem{novak2021quasi}
Nejc Novak, Oraib Al-Ketan, Lovre Krstulovi{\'c}-Opara, Reza Rowshan, Rashid K~Abu Al-Rub, Matej Vesenjak, and Zoran Ren.
\newblock Quasi-static and dynamic compressive behaviour of sheet tpms cellular structures.
\newblock {\em Composite structures}, 266:113801, 2021.

\bibitem{maheswaran2024mitigating}
B~Maheswaran, K~Chawla, and R~Thevamaran.
\newblock Mitigating oblique impacts by unraveling of buckled carbon nanotubes in helmet liners.
\newblock {\em Experimental Mechanics}, 64(2):197--209, 2024.

\bibitem{yu2019investigation}
Shixiang Yu, Jinxing Sun, and Jiaming Bai.
\newblock Investigation of functionally graded tpms structures fabricated by additive manufacturing.
\newblock {\em Materials \& Design}, 182:108021, 2019.

\bibitem{koumlis2019strain}
S~Koumlis and L~Lamberson.
\newblock Strain rate dependent compressive response of open cell polyurethane foam.
\newblock {\em Experimental mechanics}, 59:1087--1103, 2019.

\bibitem{lattanzi2015dynamic}
Ludovica Lattanzi, Ramathasan Thevamaran, Luigi De~Nardo, and Chiara Daraio.
\newblock Dynamic behavior of vertically aligned carbon nanotube foams with patterned microstructure.
\newblock {\em Advanced Engineering Materials}, 17(10):1470--1479, 2015.

\bibitem{jung2018can}
Yeonsu Jung, Young~Shik Cho, Jae~Won Lee, Jun~Young Oh, and Chong~Rae Park.
\newblock How can we make carbon nanotube yarn stronger?
\newblock {\em Composites Science and Technology}, 166:95--108, 2018.

\bibitem{pathak2009viscoelasticity}
Siddhartha Pathak, Z~Goknur Cambaz, Surya~R Kalidindi, J~Gregory Swadener, and Yury Gogotsi.
\newblock Viscoelasticity and high buckling stress of dense carbon nanotube brushes.
\newblock {\em Carbon}, 47(8):1969--1976, 2009.

\end{thebibliography}

\end{document}